\documentclass[acmsmall]{acmart}

\AtBeginDocument{
  \providecommand\BibTeX{{
    \normalfont B\kern-0.5em{\scshape i\kern-0.25em b}\kern-0.8em\TeX}}}

\setcopyright{acmcopyright}
\acmJournal{TOIT}
\acmYear{2021} \acmVolume{1} \acmNumber{1} \acmArticle{1} \acmMonth{1} \acmPrice{15.00}\acmDOI{10.1145/3450752}

\usepackage{textcomp}
\usepackage{multirow}
\usepackage{array}
\usepackage[ruled]{algorithm2e}

\definecolor{xred}{RGB}{255,0,0}
\definecolor{xblue}{RGB}{68,114,196}
\definecolor{xgreen}{RGB}{0,176,80}
\definecolor{xpurple}{RGB}{112,48,160}
\definecolor{xorange}{RGB}{237,125,49}

\begin{document}

\title{A Security Cost Modelling Framework for Cyber-Physical Systems}

\author{Igor Ivki\'c}
\affiliation{
	\institution{Lancaster University}
	\city{Lancaster}
	\country{UK}}
\email{i.ivkic@lancaster.ac.uk}

\author{Patrizia Sailer}
\affiliation{
	\institution{University of Applied Sciences Burgenland}
	\city{Eisenstadt}
	\country{AT}}
\email{patrizia.sailer@forschung-burgenland.at}

\author{Antonios Gouglidis}
\affiliation{
	\institution{Lancaster University}
	\city{Lancaster}
	\country{UK}}
\email{a.gouglidis@lancaster.ac.uk}

\author{Andreas Mauthe}
\affiliation{
	\institution{University of Koblenz-Landau}
	\city{Germany}
	\country{DE}}
\email{mauthe@uni-koblenz.de}

\author{Markus Tauber}
\affiliation{
	\institution{University of Applied Sciences Burgenland}
	\city{Eisenstadt}
	\country{AT}}
\email{markus.tauber@fh-burgenland.at}

\renewcommand{\shortauthors}{Ivki\'c, et al.}

\begin{abstract}
	Cyber-Physical Systems (CPS) are formed through interconnected components capable of computation, communication, sensing and changing the physical world. The development of these systems poses a significant challenge since they have to be designed in a way to ensure cyber-security without impacting their performance. This article presents the Security Cost Modelling Framework (SCMF) and shows supported by an experimental study how it can be used to measure, normalise and aggregate the overall performance of a CPS. Unlike previous studies, our approach uses different metrics to measure the overall performance of a CPS and provides a methodology for normalising the measurement results of different units to a common \textit{Cost Unit}. Moreover, we show how the \textit{Security Costs} can be extracted from the overall performance measurements which allows to quantify the overhead imposed by performing security-related tasks. Furthermore, we describe the architecture of our experimental testbed and demonstrate the applicability of SCMF in an experimental study. Our results show that measuring the overall performance and extracting the security costs using SCMF can serve as basis to redesign interactions to achieve the same overall goal at less costs.
\end{abstract}




\maketitle

\section{Introduction}\label{section:introduction}

A Cyber-Physical System (CPS) consists of components capable of computation, communication, sensing and changing the physical world \cite{ref24,ref26}. On their own these components often have just enough computational resources (resource constrained devices) for a specific task, e.g. controlling a sensor or an actuator. The so-called Internet of Things (IoT) can be considered as the backbone of a CPS that interconnects these constrained devices enabling them to form complex networks of embedded systems \cite{ref23}. A CPS consisting of interconnected components with sensing and actuating functionalities enables the conjunction of the cyber-verse and the physical world \cite{ref44}. Hence, this opens up a wide range of application opportunities in various domains \cite{ref48} but also poses new challenges regarding performance and security.

The main challenge regarding performance and security is to design CPS in a way to be as secure as possible without impacting the overall performance. Especially when a CPS is integrated in critical infrastructures \cite{ref28} the importance of cyber security countermeasures cannot be stressed enough. For instance, a successful cyber-attack launched at a CPS can cause disruptions transcending the cyber realm and lead to damages in the physical world \cite{ref29}. However, the practice of providing security always produces additional overhead which might lead to a compromise between performance and security especially with CPS consisting of constrained devices. Ultimately, to avoid potential performance issues requires design approaches that allow to run the interactions of a CPS efficiently while keeping a high level of security.

This brings the challenge to determine which approach is the most efficient if there is more than one potential solution. An interaction of a CPS is usually designed for a specific application purpose (main functionality) which involves a number of components that together perform a set of tasks. These tasks are directly related to the main functionality of the CPS since their execution contributes towards meeting the purpose of the application. Besides performing these \textit{functional} tasks, additional tasks are needed in order to provide security while the interaction is executed. For instance, two interacting components could perform the additional tasks of \textit{encrypting} and \textit{decrypting} messages to protect the communication from \textit{man-in-the-middle} attacks. These additional \textit{security-related} tasks are necessary to provide a specific security countermeasure in addition to the other \textit{function} tasks. To design efficient and secure interactions requires measuring the \textit{function} and \textit{security-related} tasks and aggregate them to calculate the overall performance of interactions. 

In this article the Security Cost Modelling Framework (SCMF) is introduced capable of measuring, normalising and aggregating \textit{Functional Costs} and \textit{Security Costs} of CPS. \textit{Functional Costs} refer to the measurement results of all performed \textit{functional} tasks at runtime. \textit{Security Costs} refer to the measurement results of all performed \textit{security-related} tasks at runtime. First, we present the \textit{Onion Layer Model} to formally describe interactions, their participating components and performed tasks. Next, we present four \textit{Metric Types} and show how the raw data measurement results of each type can be calculated and weighted. This allows us to use many different metrics to measure the performance of executed \textit{functional} and \textit{security-related} tasks at runtime. Furthermore, we present a methodology to normalise \cite{ref49} and aggregate these measurement results with different units to calculate the \textit{Total Costs} of CPS, where \textit{Total Costs} refer to the measurement results of all performed tasks at runtime (\textit{functional} and \textit{security-related}).

The main objective of this article is to demonstrate how the SCMF works by measuring the \textit{Functional Costs} and \textit{Security Costs} of a real CPS. This is done through an experimental study. To achieve this we present an overall use case (\textit{Closed-Loop Temperature Control}) and show two different approaches (\textit{Use Case 1}, \textit{Use Case 2}) of implementing it. In addition we describe the minimum architectural building blocks needed to setup a CPS that runs both implemented approaches of the use case (testbed). Furthermore, we describe all necessary tools to measure the performance of these two approaches at runtime. Finally, we measure, normalise, weigh, aggregate and compare the \textit{Total Costs} of both use cases and demonstrate how the SCMF can be used to evaluate which of the two solutions is performing more efficiently.

The central research questions of this article are: (i) How can the \textit{Functional Costs} and \textit{Security Costs} be measured and modelled? (ii) What \textit{Metric Types} can be used to capture the performance of interactions at runtime? (iii) How can measurement results with different units be normalised to create a generic \textit{Cost Unit}? (iv) How can \textit{Functional Costs} and \textit{Security Costs} be aggregated to calculate the \textit{Total Costs} of an interaction?

Our approach is a new proposal for measuring the \textit{Functional Costs} and \textit{Security Costs} of CPS by quantifying the resulting overhead of executed \textit{functional} and \textit{security-related} tasks. Even though there are many approaches for measuring the performance of computer systems, they either use a single metric (e.g. execution time) or only compare the results of the same \textit{Metric Types}. To the best of our knowledge there is no other approach for measuring a computing node using different metrics that are then normalised to produce a single number with a generic unit (\textit{Cost Unit}). Additionally, our proposal contains a description of the necessary architectural elements to set up a CPS consisting of distributed computing nodes and measure their \textit{Functional Costs} and \textit{Security Costs}. Furthermore, we explain how the proposed SCMF can be applied to measure all \textit{functional} and \textit{security-related} tasks using different metrics, normalise the results and aggregate them using the \textit{Onion Layer Model}. Since CPS are formed by interconnected components, the SCMF can be used at \textit{Design Time} to implement interactions, while preserving security and optimally enhancing resource usage.

\vspace{5pt}
The remainder of the article is organised as follows: In Section \ref{section:related_work}, we discuss related work. Subsequently, in Section \ref{section:model}, we present the \textit{Onion Layer Model} for describing \textit{Functional Costs} and \textit{Security Costs} in CPS, and identify four different \textit{Metric Types} to measure them. Additionally, we show how measurements with different units can be normalised and aggregated using the presented model. In Section \ref{section:architecture}, we describe the architecture that we used to demonstrate \textit{Functional Cost} and \textit{Security Cost} measurements in an implemented CPS. Then, in Section \ref{section:experiment}, we present two different possible solutions of how an interaction within the CPS could be implemented and evaluate their resulting \textit{Functional Costs} and \textit{Security Costs} in an experimental study. In Section \ref{section:discussion}, we discuss the importance of measuring the \textit{Functional Costs} and \textit{Security Costs} by referring to the results of the experimental study, and point out future applications towards digital twins and performance optimisation. Finally, in Section \ref{section:conclusion}, we conclude our work.

\section{Background and Related Work}\label{section:related_work}
In this section, we highlight the differences between existing work and ours and elaborate how the proposed SCMF fills the gap in the related work. The literature review covers measuring and modelling computational performance in general; existing performance and security metrics; and, predictive models based on past measurements. Furthermore, we also discuss related work regarding theoretical backgrounds and security concerns in Industry 4.0 including CPS and IoT-Frameworks.

\subsection{Background}

This article builds on three publications from our previous work that lay the foundation towards developing the SCMF. The initial idea regarding measuring \textit{Security Costs} was presented in \cite{ref37}, where we argued that providing security always produces an additional overhead that could be measured and analysed. However, this first publication mainly focused on how much money and time has to be spent to buy and implement security controls to design secure applications. This focus was redirected in the following publication \cite{ref38} by introducing the idea of formally describing \textit{Security Costs} of interacting components and their executed \textit{security-related} tasks in CPS. In this second publication we presented the idea of the \textit{Onion Layer Model} on a mainly theoretical level without explaining how it could be applied in practice. Therefore, in our third publication \cite{ref39}, we extended the previous paper by describing how the model could be used to measure \textit{Security Costs} of CPS at runtime. In this regard, we argued that two mechanisms are required, where one monitors which components interact with each other, while the other measures their performed tasks. Even though, this paper showed how the model could be used at runtime, it again only described the two mechanisms in theory. Furthermore, the second and third publication did not provide any answers regarding how measurement results with different units could be normalised and aggregated. 

In this article, we intend to go a step further and close the gaps of all three previous publications. In this regard, we present a framework (SCMF) and show in an experimental study how it can be used to measure, normalise and aggregate the overall performance (\textit{Total Costs}) of CPS. Moreover, we explain that the \textit{Total Costs} are made up from measuring the performance of \textit{functional} tasks (\textit{Functional Costs}) and \textit{security-related} tasks (\textit{Security Costs}). Therefore, we first present our model including four different \textit{Metric Types} and describe how each of them is used differently when measuring these tasks. Based on these \textit{Metric Types}, we then show how to calculate weights and why it is necessary to weight raw data measurement results. Next, we extend the \textit{Onion Layer Model} by providing a methodology for normalising measurement results with different units. Moreover, we present an algorithm that uses raw data measurement results as an input and returns the calculated \textit{Functional Costs} and \textit{Security Costs} as an output. In addition to that we demonstrate in an experimental study, how a CPS can be set up and its \textit{Functional Costs} and \textit{Security Costs} measured, normalised and aggregated including a description of our experimental setup (testbed).

\subsection{Economics of Information Security}

The literature review has resulted in identifying the discipline of \textit{Economics of Information Security} as the high-level research field which (according to \cite{ref41}) can be categorised in the following sub-disciplines of \textit{Economics of Privacy}, \textit{The Information Security Business}, \textit{Economics of Vulnerabilities}, \textit{Measuring Electronic Crime}, and \textit{Information Security Regulation}. In the early 2000s the \textit{Economics of Information Security} emerged as a thriving and fast-moving research discipline, where the interest initially came from the banking industry \cite{ref01}. Even though banks spent more on security to prevent card fraud, the number of attacks still continued to increase. Several research streams have been developing, particularly investigation into \textit{Measuring the Cost of Cyber-Crime} \cite{ref02}, \textit{Review on the Scale and Nature of Cyber Crime} \cite{ref03}, \textit{Examination of Impact of Notice and Take-Down} \cite{ref04,ref05}, and \textit{Analysis on various Cyber-Crime Attacks} \cite{ref06,ref07,ref08}. However, most of the presented approaches and frameworks provide methodologies for estimating/measuring the size of the investment when buying hardware/software (including configuration) for providing cyber security for an existing system. Other approaches and frameworks focus on evaluating how much harm (or cost) a specific cyber-attack caused and how much it costs to repair the resulting damage. To the best of our knowledge, in the field of \textit{Economics of Information Security} there are currently no approaches for measuring, how much security costs during the operation of a system. In other words, how high is the overhead for providing security for a specific system at runtime. This question is aggravated even more when the system is a CPS, where a cyber-attack might cause disastrous consequences in both the cyber-verse and the physical world \cite{ref43}.

\subsection{Measuring Cyber Security without Security Costs-Relatedness}
In other related work, there are many studies that focus on measuring cyber security, but they do so without referring to the resulting costs of providing security. In addition to that some of the presented approaches are often limited by the usage of a single metric. For instance, in \cite{ref09,ref10} the authors show how the performance of a process can be measured, while the focus in \cite{ref11,ref12,ref13,ref14} is on evaluating the energy consumption. A summary of related work regarding security metrics has been provided by Yee in \cite{ref21}. He first establishes the argument that many security metrics exist, but most of them are ineffective and not meaningful. Next, he provides a definition of a "good" and a "bad" metric and explains the difference between "traditional" and "scientifically based" security metrics. Finally, Yee presents his literature search on security metrics which is based on various frameworks. Other related work use methods and frameworks to evaluate how secure e.g. a system is \cite{ref15,ref16,ref17,ref18,ref19,ref20}. In other words, these solutions focus on evaluating whether a security control has been implemented or not and how much effort it takes (or how much it costs) to do so. This approach of measuring the costs of implementing security controls using money as a metric is similar to our first publication in \cite{ref37}. The work of \citeauthor{ref50} \cite{ref50} comes closest to our presented approach in this article, where they used different metrics to measure the \textit{Security Costs} of the Hypertext Transfer Protocol Secure (HTTPS). The major difference compared to our work is that \citeauthor{ref50} measured the HTTP and then compare the results to the measurements of the HTTPS. These measurement results were compared (metric by metric) independently to each other to shed some lights on the impacts on latency, data consumption and battery life for clients. In this article we take one step further towards aggregating \textit{Functional Costs} and \textit{Security Costs} of a CPS, by using many different metrics for measuring and a normalisation method to create a common \textit{Cost Unit}.

\subsection{Engineering Challenges and Security Concerns in Industry 4.0}
The vast majority of research related to Industry 4.0 and CPS currently focuses more on general challenges \cite{ref22,ref23}, design principles \cite{ref24,ref25,ref40}, or engineering \cite{ref26,ref27}. However, in \cite{ref28} the authors give an overview of the security concerns in CPS, identify challenges and summarise countermeasures. Rajkumar, De Niz and Klein \cite{ref29} demonstrate further that the complexity of CPS requires more effort to analyse and defend it. The reason for that is the explosion of states when considering combinations of events. Additionally, they provide theoretical approaches for dealing with cyber threats and countermeasure models for CPS under attack. 

\subsection{CPS and IoT-Frameworks}
Regarding CPS and the IoT, there are various approaches, platforms and frameworks supporting the Industry 4.0 movement. \citeauthor{ref30} \cite{ref30} summarise commercially available IoT frameworks including the IoTivity framework \cite{ref31}, the IPSO Alliance framework \cite{ref32}, the Light Weight Machine to Machine (LWM2M) framework \cite{ref33}, the AllJoyn framework \cite{ref34} and the Smart Energy Profile 2.0 (SEP2.0) \cite{ref35}. Most of the cloud-based frameworks follow a data-driven architecture in which all involved IoT-components are connected to a global cloud using one Service Oriented Architecture (SOA) protocol. The Arrowhead Framework \cite{ref36}, on the contrary, follows an event-driven approach, in which a \textit{local cloud} is governed through the use of core systems for registering and discovering services, authorisation and orchestration. Since everything within an Arrowhead \textit{local cloud} is a service, new supporting systems can be developed and added to the already existing ones. This enables implementing and deploying any use case independent of complexity and number of components running supporting systems. For this reason, we chose to deploy our overall use case and draw the system boundaries of the deployed CPS by using the Arrowhead Framework.

\subsection{Performance Measurement using an Agent-Based Approach}
Another reason for using the Arrowhead Framework was that it is an open source project implemented in Java \cite{ref62}. This allows us to instrument the code of all components within the deployed Arrowhead \textit{local cloud} and measure the \textit{Functional Costs} and the \textit{Security Costs} of all interactions. To overcome the challenge of instrumenting each line of code manually \cite{ref51} we used Pinpoint \cite{ref47}, which is an agent-based Application Performance Management (APM) tool modelled after Google’s Dapper \cite{ref52}. Pinpoint uses agents to instrument the bytecode of a Java application at runtime and capture performance data using different built-in metrics. A further advantage: Pinpoint was built for large-scale distributed systems and has a minimal impact on the overall performance\cite{ref53,ref54}.

\subsection{Summary}

As discussed in this section and summarised in Table \ref{tab:rwsum}, the related work focuses either on,
\begin{itemize}
	\item measuring the costs of buying hardware and software for protecting computer systems, or
	\item calculating the damage caused by specific cyber-attacks, or
	\item evaluating how secure a system is, or
	\item using just a single metric to measure a specific aspect of a system, or
	\item comparing measurements results on a "metric by metric"-basis.
\end{itemize}

\vspace{5pt}
This article aims at closing the identified gaps by introducing the SCMF for quantifying \textit{Security Costs} of CPS. The proposed framework follows the idea of measuring performed tasks of each component of an interaction by using many different metrics. Since different metrics produce results with different units, the SCMF also provides a methodology for normalising these measurement results to a common \textit{Cost Unit} which then can be aggregated. In this regard, we demonstrate in an experimental study how the SCMF can be used to measure, analyse and compare the interactions in two different use cases. 

\begin{table}[H]
	\centering
	\caption{Measuring Security Costs (Summary of Literature Review)}
	\label{tab:rwsum}
	\begin{tabular}{lc}
		\toprule
		\textbf{\textit{Economics of Information Security}}&\textbf{\textit{Sources}}\\
		\midrule
		\quad Introduction to Economics and Information Security & \cite{ref01} \\
		\quad Measuring the impact and resulting costs of cybercrimes. & \cite{ref02} \\
		\quad Review on the scale and nature of cyber crime. & \cite{ref03} \\
		\quad Examination of impact of notice and take-down. & \cite{ref05,ref04} \\
		\quad Measuring the impact of various cyber-crime attacks. & \cite{ref07,ref06,ref08} \\
		\toprule
		\textbf{\textit{Measuring Cyber Security without Security Cost-Relatedness}}&\textbf{\textit{Sources}}\\
		\midrule
		\quad Summary on measurable security metrics. & \cite{ref21} \\
		\quad Measuring process performance \& complexity. & \cite{ref09,ref10} \\
		\quad Measurement of performance and energy consumption. & \cite{ref14,ref13,ref11,ref12} \\
		\quad Measuring how secure a specific system is (level of security). & \cite{ref15,ref16,ref20,ref18,ref19,ref17} \\
		\quad Measuring the impact of various cyber-crime attacks. & \cite{ref50} \\
		\toprule
		\textbf{\textit{Engineering Challenges and Security Concerns in Industry 4.0}}&\textbf{\textit{Sources}}\\
		\midrule
		\quad General challenges of Industry 4.0 and CPS & \cite{ref23,ref22} \\
		\quad Design principles and challenges of CPS. & \cite{ref24,ref40,ref25} \\
		\quad Engineering challenges regarding CPS and Industry 4.0 applications. & \cite{ref26,ref27} \\
		\quad Security concerns and analysis requirements to defend and secure CPS. & \cite{ref43,ref28,ref29} \\
		\toprule
		\textbf{\textit{CPS and IoT-Frameworks}}&\textbf{\textit{Sources}}\\
		\midrule
		\quad Review of existing frameworks supporting the Industry 4.0 movement. & \cite{ref30} \\
		\quad Data-driven IoT-Frameworks running in a global cloud. & \cite{ref31,ref32,ref33,ref34,ref35} \\
		\quad Event-driven IoT-Frameworks running in a local cloud. & \cite{ref36} \\
		\toprule
		\textbf{\textit{Performance Measurement using an Agent-Based Approach}}&\textbf{\textit{Sources}}\\
		\midrule
		\quad Bytecode instrumentation and APM-tools. & \cite{ref51,ref47,ref52,ref54,ref53} \\
		\bottomrule
	\end{tabular}
\end{table}

\section{Model}\label{section:model}

In this section, we present an approach for evaluating \textit{Functional Costs} and \textit{Security Costs} in a CPS by abstracting all interactions within the CPS using the \textit{Onion Layer Model}. By doing so we describe the functionality of the \textit{Onion Layer Model} and explain how it can be used to model the entirety of all interactions, their participating components and performed tasks within a CPS. Next, we show that measuring \textit{Functional Costs} and \textit{Security Costs} requires different measurement techniques and summarise them in four \textit{Metric Types}. Based on the \textit{Metric Types} we explain how to calculate a weight which can be used to emphasise each result more or less. Furthermore, we present a normalisation method which can be used to transform the raw data measurement results with different units (based on the usage of different metrics) to a common \textit{Cost Unit}. Finally, we present an algorithm for normalising and aggregating \textit{Functional Costs} and \textit{Security Costs}.

\subsection{Onion Layer Model}

As previously mentioned, a CPS is formed by a number of components that are capable of measuring the physical world by using sensors and changing it by controlling actuators. In addition these components are capable of computation and communication using the IoT infrastructure, where they are interconnected with each other. Together they form logical units within a CPS by participating in one or more interactions to reach a common goal or serve a specific purpose. For instance, one component could be using a sensor to measure the temperature of a room while another component uses the measured data to control an air-conditioning system. The overall goal or purpose of this interaction would be to measure and control the temperature of a physical room. To achieve this each component performs a number of different tasks which all together contribute towards reaching this common goal. For instance, a component could be using a sensor to measure the temperature of a room and sending the measured data to another component. This second component verifies whether a predefined temperature limit has been reached and decides to activate the air-conditioning system to cool down the room (if necessary). 

An interaction in the context of a CPS refers to a logical unit that is independently executed at a specific point in time and treated in a coherent and reliable way, independent of other interactions. This means that one CPS cannot have two different interactions where the same components participate in both of them at the same time and perform the exact same tasks. A redundancy like this, would go against any design principle in engineering if it was not for the purpose of duplication of critical components to increase reliability (e.g. backups or high-availability). That being said, the total number of all interactions, the participating components, and their performed tasks can be described as follows (for any set $X$, we use $\widehat{X}$ to denote its upper bound in this article):
\begin{itemize}
	\item[] $I$ = \{$i_1$,$i_2$,…,$i_{\widehat{I}}$\}, is the set of all interactions of $CPS_1$ where $\widehat{I}$ defines the upper bound of $I$
	\item[] $C$ = \{$c_1$,$c_2$,…,$c_{\widehat{C}}$\}, is the set of all components in $I$, where $\widehat{C}$ defines the upper bound of $C$
	\item[] $T$ = \{$t_1$,$t_2$,…,$t_{\widehat{T}}$\}, is the set of all tasks in $C$, where $\widehat{T}$ defines the upper bound of $T$
\end{itemize}

\vspace{5pt}
For instance, within one interaction $i_1$ there could be a set of components $C{_{i_1}}$= \{$c_1$,$c_2$,$c_3$\}, while another interaction $i_2$ has a different set of components $C_{{i_2 }}$= \{$c_1$,$c_2$,$c_3$,$c_4$\}. This means that the size of a component set depends on the number of its components within a specific interaction. In the previous example $\widehat{C}_{{i_1}}$ compared to $\widehat{C}_{{i_2}}$ has a smaller number of components ($\widehat{C}_{{i_1}}\textless\widehat{C}_{{i_2}}$). Similar to that the size of a task set $\widehat{T}$ also depends on the number of its tasks performed by a specific component. If $C_1$ performed fewer tasks in comparison to $C_2$, the task set of the first component would be smaller compared to the task set of the second component $(\widehat{T}_{{c_1}}\textless\widehat{T}_{{c_2}})$. As mentioned before, some of the performed tasks can be directly related to the main functionality, while others could be directly related to providing security. As an example, the task for measuring the temperature of a room is related to the function, while encrypting a message before it is sent directly corresponds to a \textit{security-related} task. 
To measure the performance of a task, the \textit{Onion Layer Model} suggests to use a set of different metrics $M$ = \{$m_1$,$m_2$,…,$m_{\widehat{M}}$\}, where $\widehat{M}$ defines the upper bound of $M$. Finally, the measured tasks need to be categorised in \textit{functional} and \textit{security-related} tasks to be able to extract the \textit{Security Costs} from the \textit{Total Costs}. As shown in Fig. \ref{fig:Fig1}, the \textit{Onion Layer Model} describes a CPS including all interactions, the participating components, their performed tasks, and all metrics used for measuring the costs:
\vspace{-10pt}
\begin{figure}[H]
	\centering
	\includegraphics[width=9.2cm]{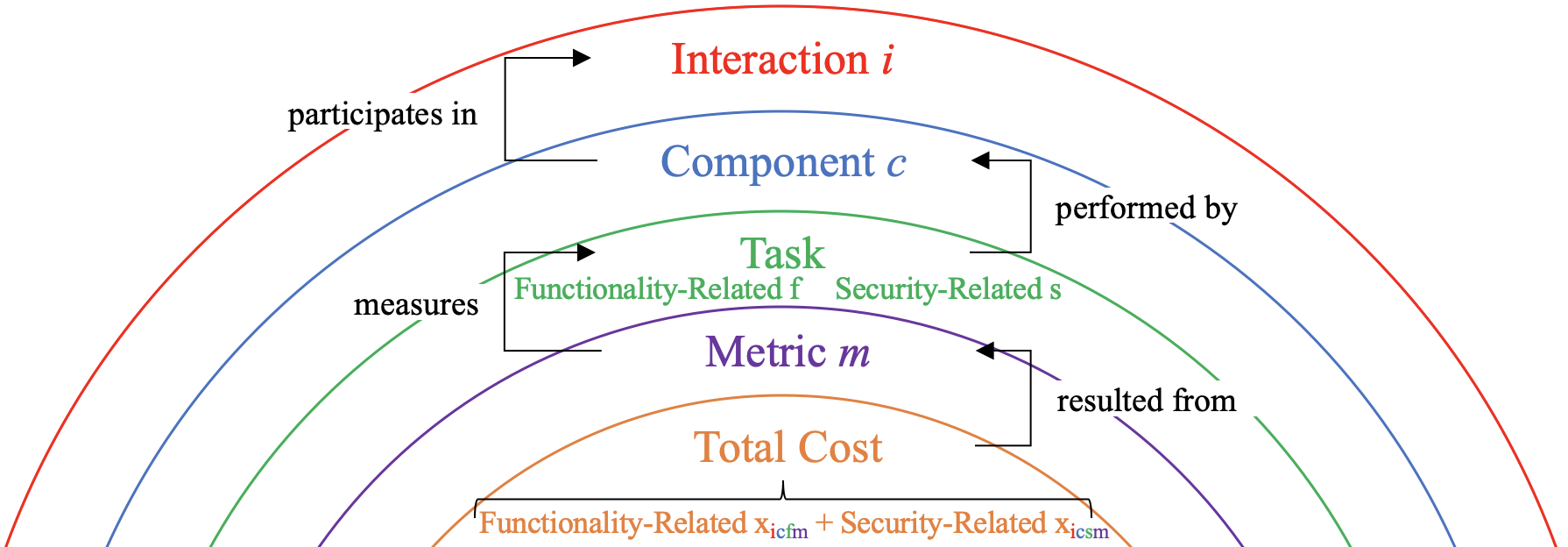}
	\caption{Onion Layer Model for Modelling \textit{Functional Costs} and \textit{Security Costs} (adapted from \citeauthor{ref39}, 2019).}
	\Description{Onion Layer Model.}
	\label{fig:Fig1}
\end{figure}

\vspace{-5pt}
The \textit{Onion Layer Model} allows describing the entirety of interactions within a specific CPS and enables modelling the \textit{Functional Costs} and \textit{Security Costs} resulting from measuring \textit{functional} and \textit{security-related} tasks. In other words, by using metric $m_1$ to measure all performed tasks the measurement results of \textit{functional} tasks in comparison to the \textit{security-related} tasks can be put in perspective. As shown in (\ref{eq:1}), the \textit{Onion Layer Model} enables modelling the \textit{Total Costs} of an interaction by summing up the \textit{Functional Costs } and the \textit{Security Costs} of a CPS:
\begin{equation}\label{eq:1}
	\centering
	\begin{split}
		TotalCosts_{CPS_1} &= FunctionalCosts_{CPS_1} + SecurityCosts_{CPS_1} \\
		FunctionalCosts_{CPS_1} &= \textcolor{xred}{\sum_{i=1}^{\widehat{I}_{CPS_1}}}\textcolor{xblue}{\sum_{c=1}^{\widehat{C}_i}}\textcolor{xgreen}{\sum_{f=1}^{\widehat{F}_c}}\textcolor{xpurple}{\sum_{m=1}^{\widehat{M}_f}}(\textcolor{xorange}{\dot{x}_{\textcolor{xred}{i}\textcolor{xblue}{c}\textcolor{xgreen}{f}\textcolor{xpurple}{m}}} * w_{MT}) \\
		SecurityCosts_{CPS_1} &= \textcolor{xred}{\sum_{i=1}^{\widehat{I}_{CPS_1}}}\textcolor{xblue}{\sum_{c=1}^{\widehat{C}_i}}\textcolor{xgreen}{\sum_{s=1}^{\widehat{S}_c}}\textcolor{xpurple}{\sum_{m=1}^{\widehat{M}_s}}(\textcolor{xorange}{\dot{x}_{\textcolor{xred}{i}\textcolor{xblue}{c}\textcolor{xgreen}{s}\textcolor{xpurple}{m}}} * w_{MT})
	\end{split}
\end{equation}

\vspace{-10pt}
\begin{table}[H]
	\centering
	\begin{tabular}{cl}
		\multirow{2}{*}{$TotalCosts_{CPS_1}$} & represents the sum of the measured, normalised and aggregated\\
		& \textit{Functional Costs} and \textit{Security Costs} of $CPS_1$\\
		\rule{0pt}{15pt}$\textcolor{xred}{\sum_{i=1}^{\widehat{I}_{CPS_1}}}$ & represents the set of all interactions of $CPS_1$  \\
		\rule{0pt}{15pt}$\textcolor{xblue}{\sum_{c=1}^{\widehat{C}_i}}$ & represents the set of all components of interaction $i$ \\
		\rule{0pt}{15pt}\multirow{3}{*}{$\textcolor{xgreen}{\sum_{f=1}^{\widehat{F}_c}}$,  $\textcolor{xgreen}{\sum_{s=1}^{\widehat{S}_c}}$} & represents the set of all \textit{functional} and \textit{security-related} tasks performed by\\
		& component c, where $\widehat{F}_c$ is the set of all \textit{functional} tasks and $\widehat{S}_c$ is the set of \\
		& all \textit{security-related} tasks, but both are the union set of all tasks $\widehat{T}_c = \widehat{F}_c \cup \widehat{S}_c$\\
		\rule{0pt}{15pt}$\textcolor{xpurple}{\sum_{m=1}^{\widehat{M}_f}}$, $\textcolor{xpurple}{\sum_{m=1}^{\widehat{M}_s}}$ & represents the set of all metrics used to measure all performed tasks \\
		\rule{0pt}{15pt}\multirow{3}{*}{$\textcolor{xorange}{\dot{x}}_{\textcolor{xred}{i}\textcolor{xblue}{c}\textcolor{xgreen}{f}\textcolor{xpurple}{m}}$, $\textcolor{xorange}{\dot{x}}_{\textcolor{xred}{i}\textcolor{xblue}{c}\textcolor{xgreen}{s}\textcolor{xpurple}{m}}$} & $\dot{x}$ is the normalised value of $x$, where $x$ is the raw data measurement result of\\
		& metric $m$ used to measure task $f$ (\textit{functional}) or task $s$ (\textit{security-related})\\ 
		& performed by component $c$, during interaction $i$ \\
		\rule{0pt}{15pt}\multirow{2}{*}{$w_{MT}$} & represents weight \textit{w} of Metric Type \textit{$MT$} which can be used to emphasise \\
		& a normalised result $\dot{x}$, where \textit{$MT$} $\in$ \{$MT_1$, $MT_2$, $MT_3$\}
	\end{tabular}
	\label{tab:eq1}
\end{table}

As shown in (\ref{eq:1}), both the \textit{Functional Costs} and \textit{Security Costs} of $CPS_1$ are calculated by four different sums. The first sum  $\textcolor{xred}{\sum_{i=1}^{\widehat{I}_{CPS_1}}}$ represents all existing interactions of $CPS_1$, while the second sum $\textcolor{xblue}{\sum_{c=1}^{\widehat{C}_i}}$ holds all components within one interaction \textit{i}. Depending on the tasks type, the next sum ($\textcolor{xgreen}{\sum_{f=1}^{\widehat{F}_c}}$, $\textcolor{xgreen}{\sum_{s=1}^{\widehat{S}_c}}$) either all \textit{functional} or \textit{security-related} tasks performed by one component \textit{c}. In this regard, both sets together form the union set of all performed tasks of interaction $i$ ($\widehat{T}_c = \widehat{F}_c \cup \widehat{S}_c$). Finally, the last sum ($\textcolor{xpurple}{\sum_{m=1}^{\widehat{M}_f}}$, $\textcolor{xpurple}{\sum_{m=1}^{\widehat{M}_s}}$ ) adds up all metrics which have been used to measure the costs of one task \textit{f} or \textit{s}. As already mentioned, whenever different metrics are being used for measurements it leads to having many results with different units. Thus, to be able to aggregate all measurement results using the \textit{Onion Layer Model} a normalisation methodology has to be used. The formal description $\dot{x}_{\textcolor{xred}{i}\textcolor{xblue}{c}\textcolor{xgreen}{f}\textcolor{xpurple}{m}}$ or $\dot{x}_{\textcolor{xred}{i}\textcolor{xblue}{c}\textcolor{xgreen}{s}\textcolor{xpurple}{m}}$ specifies the normalised measurement result of metric $m$ used to measure task $u$ or $s$ which has been performed by component $c$ that participated in interaction $i$. Thus, this is an extension of the \textit{Onion Layer Model} in \cite{ref38,ref39} to enable aggregating these measurement results with different units by normalising them to a generic \textit{Cost Unit}.

\subsection{Metric Types and Weight Calculation}

As described by the \textit{Onion Layer Model} in (\ref{eq:1}), the resulting \textit{Total Costs} of an interaction are a reflection of performance measurements of interacting components and their performed tasks. From a metrological perspective, there are many ways of measuring the performance of a specific task depending on the used metric and how the measurement was carried out. In principle, there are three basic characteristics of a computing node that are typically measured. The first characteristic refers to counting how often an event occurs, the second measures the duration of some time interval and the third considers the size of some parameter \cite{ref55}. At first glance the first characteristic might seem promising when used as a metric to simply count the number of tasks within an interaction. Even though it is not a big challenge to measure the number of executed tasks, the result of such a measurement might give a false impression of actual costs. For instance, an interaction could be designed in a way that it executes three \textit{functional} and two \textit{security-related} tasks, meaning a total of five tasks. More precisely, this would mean that 40\% of all executed tasks of this interaction are directly related to security. However, if we used the second characteristic to measure how long it takes to execute each task it might lead to a completely different result. For example, the results of the second measurement could be that the execution of each \textit{functional} task takes 3 seconds, while the \textit{security-related} tasks take 0.5 seconds each. In total, the second result shows that the entire interaction takes 10 seconds, where the \textit{functional} tasks take 9 seconds compared to the \textit{security-related} tasks with 1 second. This means that only 10\% of all executed tasks are directly related to security which shows a different picture to the previously measured 40\% when counting tasks. Based on the basic characteristics and the above reflections, we derived the following four \textit{Metric Types} that can be used to measure the performance of interactions:
\vspace{-11pt}
\begin{figure}[H]
	\centering
	\includegraphics[width=7cm]{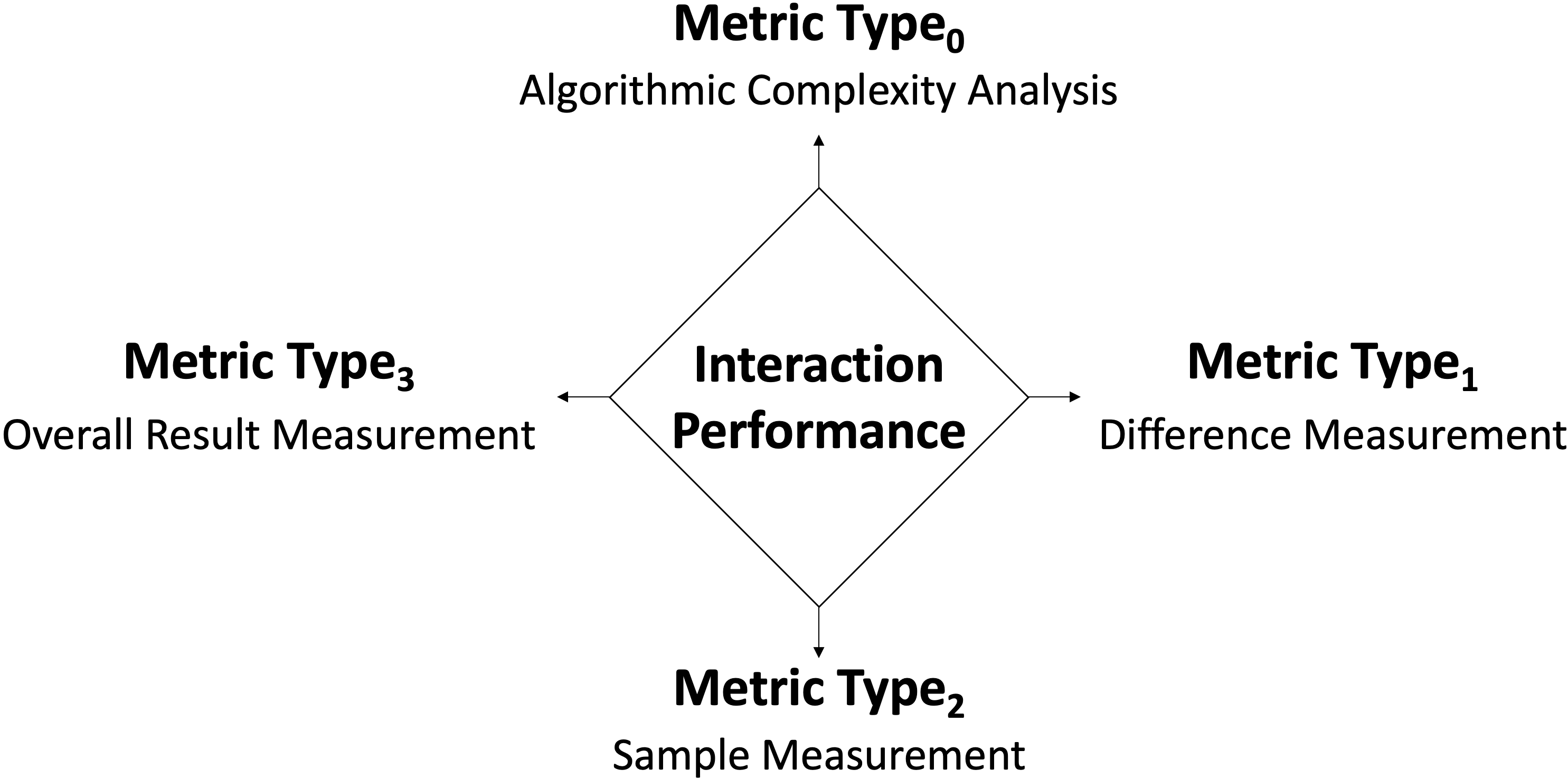}
	\caption{Metric Types for Measuring the Performance of Interactions.}
	\Description{Metric Types.}
	\label{fig:Fig2}
\end{figure}

\subsubsection{$Metric Type_0$ (Algorithmic Complexity Analysis)}

Algorithm analysis aims at providing a rough estimate of the resources an algorithm may need to solve a computational problem. Ultimately, these estimates are then used to determine and compare the efficiencies of different algorithms \cite{ref56}. Due to the \textit{Halting Problem} analysing algorithmic complexity cannot be done automatically (or at run-time). The \textit{Halting Problem} describes, based on a description of an arbitrary computer program and an input, whether the program will stop running (halt) or continue to run forever. Furthermore, \citeauthor{ref57} \cite{ref57} proved that a general algorithm to solve this problem for all possible program-input pairs cannot exist. Thus, this means that it is impossible to implement a program that analyses the complexity of a given algorithm, since it cannot be determined, whether the algorithm will halt. Knowing this, we propose to use the Big-O notation to manually estimate how complex a specific algorithm of a component is. Using the Big-O notation enables to classify the complexity of a components algorithm. Consequently, the component with the better performing algorithm costs less in comparison to a component with a more complex algorithm. Even though $Metric Type_0$ can only be measured manually, it can be significant to know how performant a specific algorithm is. For instance, if two components produce the same \textit{Total Costs}, but one of them has a better performing algorithm according to the Big-O estimate, they can still be compared. In that case the results of the Big-O estimation can be used as a constant that is added to the overall result of the \textit{Onion Layer Model}. Furthermore, since the Big-O can be used to estimate the \textit{Algorithmic Time Complexity} and the \textit{Algorithmic Space Complexity}, a constant representing each of them can be added to the overall result of the \textit{Onion Layer Model}. The following shows the algorithmic complexities (Big-O) ordered from low complexity $f_{MT_0}(1) $ to high complexity $f_{MT_0}(n!)$:
\begin{equation}\label{eq:2}
	f_{MT_0}(1) < f_{MT_0}(\log{}n) < f_{MT_0}(n) < f_{MT_0}(n\log{}n) < f_{MT_0}(n^2) < f_{MT_0}(2^n) < f_{MT_0}(n!)
\end{equation}

\subsubsection{$Metric Type_1$ (Difference Measurement)}

This metric type aims at capturing specific evaluation criteria before the execution of a task and right after it to be able to calculate the difference between those two captures. The most commonly used representative of this type is the \textit{duration} or \textit{time required to execute} a given application program. The basic technique for measuring how long it takes to perform a task is to set a timestamp before and after it, and then calculate the difference between both timestamps. Another example for this metric type would be to measure how a property changes before and after the execution of a certain task. For instance, the character length of a message is different before and after using an encryption method. The same effect is given with different communication protocols, since they add extra bytes to a given message depending on the implementation of the protocol. All in all, determining the difference between two measurement points is a key candidate for measuring the performance of interactions at runtime. The following equation shows how this metric type can be used to calculate the difference between two capture points for all tasks performed by a specific component $c$:
\vspace{-5pt}
\begin{equation}\label{eq:3}
	X_{MT_1c} = \sum_{t=1}^{\widehat{T}_{c}}  x_{t_{end}} - x_{t_{start}}\text{, where}
\end{equation}

\vspace{-10pt}
\begin{table}[H]
	\centering
	\begin{tabular}{cl}
		\multirow{2}{*}{$X_{MT_1c}$} & represents the calculated raw data measurement value of $Metric Type_1$ ($MT_1$) for all\\ 
		& \textit{functional} or \textit{security-related} tasks performed by component $c$ \\
		\rule{0pt}{15pt}\multirow{2}{*}{$\sum_{t=1}^{\widehat{T}_{c}}$} & represents the sum of all difference measurement values of all \textit{functional} or\\ 
		& \textit{security-related} tasks \textit{t} performed by component $c$ \\
		\rule{0pt}{15pt}$x_{t_{start}}$ & represents the capture before task $t$ (\textit{functional} or \textit{security-related}) starts \\
		\rule{0pt}{15pt}$x_{t_{end}}$ & represents the capture after task $t$ (\textit{functional} or \textit{security-related}) ended
	\end{tabular}
	\label{tab:eq3}
\end{table}

\subsubsection{$Metric Type_2$ (Sample Measurement)}

This metric type requires to repeatedly create snapshots during the course of executing a specific task. By doing so, many different snapshots are being created which capture the progression of the performance during the execution time. In this regard, measuring performance refers mainly to the usage of computing resources like the Central Processing Unit (CPU), Random Access Memory (RAM), disk memory, and so on. In general, it does not make any sense to measure how much CPU has been used before and after the execution of a task. The reason for that is that the CPU utilisation at these two points in time would either be near zero or not represent the actual usage. On the contrary, measuring the usage of computing resources has to be constantly monitored to produce more meaningful and accurate results. In a more metaphorical sense, evaluating the utilisation of computing resources can be compared to measuring the heart beat of a jogging person. If there are only two heartbeat measurements, one before running and one right after that, the results would neither have any meaning nor be accurate enough. However, using the sample measurement type to repeatedly measure the heartbeat during the run would produce results that reflect the performance of the jogger. Since each task can have a different set of sample measurements $m_i$, we calculate the mean of all sample measurements of all tasks $t$ performed by component $c$ as shown in the following equation:
\vspace{-5pt}
\begin{equation}\label{eq:4}X_{MT_2c} = \frac{\sum_{t=1}^{\widehat{T}_c}\sum_{n=1}^{m_i}x_{tn}}{\sum_{i=i}^{j}m_i}\text{, where}\end{equation}

\vspace{-15pt}
\begin{table}[H]
	\centering
	\begin{tabular}{cl}
		\multirow{2}{*}{$X_{MT_2c_i}$} & represents the mean of $Metric Type_2$ ($MT_2$) for all sample measurement values\\
		& for task $t$ (\textit{functional} or \textit{security-related}) performed by component $c$\\
		\rule{0pt}{15pt}\multirow{2}{*}{$\sum_{t=1}^{\widehat{T}_c}\sum_{n=1}^{m_i}x_{tn}$} & represents the total sum of all sample measurement values $x_{tn}$ of all \textit{functional} \\
		& or \textit{security-related} tasks performed by component $c$\\
		\rule{0pt}{15pt}$\sum_{n=1}^{m_i}x_{tn}$ & represents the sample size of all measured tasks
	\end{tabular}
	\label{tab:eq4}
\end{table}

\subsubsection{$Metric Type_3$ (Overall Result Measurement)}

The Type-3 metric type refers to all measurements that cannot or are difficult to pin down to a specific task. In some cases, it will be hard to determine which task is responsible for a specific measurement value at a specific point in time. Even though measuring the power consumption in CPS is very important, it is hard to tell which part of an executed program code consumed a specific amount of energy. To solve this problem, the total energy consumption for all tasks performed by a component is measured and than multiplied by the duration of performed tasks (\textit{functional} or \textit{security-related}). The following equation shows how to calculate the costs of $Metric Type_3$:
\vspace{-5pt}
\begin{equation}\label{eq:5}X_{MT_3c} = x_{total} * X_{MT_1c}\text{, where}\end{equation}

\vspace{-5pt}
\begin{table}[H]
	\centering
	\begin{tabular}{cl}
		\multirow{2}{*}{$X_{MT_3c}$} & represents the calculated overall result of $Metric Type_3$ ($MT_3$) for all \textit{functional} or\\
		& \textit{security-related} tasks performed by component $c$\\
		\rule{0pt}{15pt}\multirow{2}{*}{$x_{total}$} & represents the total raw data measurement value of component $c$ while performing \\
		& all \textit{functional} or \textit{security-related} tasks\\
		\rule{0pt}{15pt}\multirow{2}{*}{$X_{MT_1c}$} & represents the calculated raw data measurement value of $Metric Type_1$ ($MT_1$) of all \\ 
		& \textit{functional} or \textit{security-related} tasks performed by component $c$, as described in (\ref{eq:3}) \\
	\end{tabular}
	\label{tab:eq5}
\end{table}

\subsubsection{Weight Calculation}

As suggested by the \textit{Onion Layer Model} in (\ref{eq:1}) a set of different metrics can be used to measure the performance of a specific task. Whereby, each metric can be an element belonging to $Metric Type_1$, $Metric Type_2$ or $Metric Type_3$. At the same time, each metric belonging to one of these three \textit{Metric Types} also belongs to $MetricType_0$, since the execution of a task always requires an algorithm. This distinction (which metric belongs to which \textit{Metric Type}) is important, since the weight $w$ is calculated by the number of metrics of a specific metric type. The more metrics of the same \textit{Metric Type} are used to measure a specific task the higher the weight has to be to emphasise the measurement results. For instance, a task could be measured by four metrics, where two metrics belong to $Metric Type_1$, one belongs to $Metric Type_2$ and the last is part of $Metric Type_3$. Thus, all measurement results by the metrics of $Metric Type_1$ are weighted by 0.5 ($w_{MT_1} = 0.5$), while the other results are weighted by 0.25 ($w_{MT_2} = 0.25$ and $w_{MT_3} = 0.25$). Following this example, the measurement results multiplied by $w_{MT_1}= 0.5$ are considered to have a bigger impact on the overall result compared to the other two weights ($w_{MT_2}= 0.25$ and $w_{MT_3}= 0.25$). The reason for that is that more metrics of $Metric Type_1$ have been used and therefore have more influence on the total result. In this example, the measurement results multiplied by $w_{MT_1}= 0.5 $ are considered to have a bigger impact on the overall result compared to the measurement results multiplied by $w_{MT_2}= 0.25$ and $w_{MT_3}= 0.25$. The following equation shows how the weight is calculated based on the number of metrics used for a specific \textit{Metric Type}:
\vspace{-5pt}
\begin{equation}\label{eq:6}w_{MT_j} = \frac{n_{MT_j}}{\widehat{M}_t}\text{, where}\end{equation}

\vspace{-15pt}
\begin{table}[H]
	\centering
	\begin{tabular}{cl}
		%\multicolumn{2}{c}{\vbox{\begin{equation}\label{eq:6}w_{MT_j} = \frac{n_{MT_j}}{\widehat{M}_s}\text{, where}\end{equation}}}\\
		$w_{MT_j}$ & represents the calculated weight for \textit{Metric Type} $j$ $(MT_j)$\\
		\rule{0pt}{15pt}\multirow{2}{*}{$n_{MT_j}$} & is the total number of metrics that belong to \textit{Metric Type} $j$ $(MT_j)$ and have been used \\
		& to measure task $t$ (\textit{functional} or \textit{security-related})\\
		\rule{0pt}{15pt}$\widehat{M}_t$ & is the total number of metrics used to measure task $t$ (\textit{functional} or \textit{security-related})
	\end{tabular}
	\label{tab:eq6}
\end{table}

\subsection{Normalisation}

As pointed out by \citeauthor{ref49} in \cite{ref49}, depending on the study domain there are several definitions of data normalisation. In this article, we use unity-based normalisation (\textit{MIN-MAX Normalisation} or \textit{Feature Scaling}) to bring all measurement values with different units into the range $[a,b]$. This is a generalisation to restrict the range of values between the points $a$ (lower bound) and $b$ (upper bound). For simplicity reasons, we will use the range $[a = 0,b = 1]$ in this article when normalising any measurement results. Generally speaking, as summarised by \citeauthor{ref49} \cite{ref49}, any normalisation technique that brings the measurement values into a range between two points can be used. We propose the following adapted version of \textit{MIN-MAX Normalisation}, where the MIN- and MAX-values depend on the lower and upper bounds of the \textit{Metric Type} used to measure result \textit{x}. The following equation defines how to calculate the normalised value $\dot{x}$ of the measured raw data value $x$ between the lower $(MIN_{MT_j})$ and upper $(MAX_{MT_j})$ bounds of \textit{Metric Type} $j$ $(MT_j)$ between $[a,b]$:
\vspace{-5pt}
\begin{equation}\label{eq:7}\dot{x} = a + \frac{(x - MIN_{MT_j}) * (b - a)}{MAX_{MT_j} - MIN_{MT_j}}\text{, where}\end{equation}

\vspace{-15pt}
\begin{table}[H]
	\centering
	\begin{tabular}{cl}
		$\dot{x}$ & is the normalised value\\
		\rule{0pt}{15pt}$x$ & is the measured raw data value\\
		\rule{0pt}{15pt}$MIN_{MT_j}$ & represents the lower bound of \textit{Metric Type} $j$ $(MT_j)$\\
		\rule{0pt}{15pt}$MAX_{MT_j}$ & represents the upper bound of \textit{Metric Type} $j$ $(MT_j)$\\
		\rule{0pt}{15pt}$a, b$ & represent the normalisation range where $a \leq \dot{x} \leq b$ (a = min, b = max)
	\end{tabular}
	\label{tab:eq7}
\end{table}

\subsection{Aggregation}

Following the steps of the previous sections, the performance of an interaction can be measured at runtime by different metrics belonging to one of three different \textit{Metric Types}. As a result, the functions in (\ref{eq:3}) (\ref{eq:4}) (\ref{eq:5}) can be used to calculate the raw data measurement results of the performance of each task. In addition to that $Metric Type_0$ can be used to estimate and classify (\ref{eq:2}) how well the algorithm of each component performs for a given set of $n$ tasks. Once the performance of each task has been measured, the next step is to normalise each raw data measurement result ($x$) in order to create a normalised measurement result ($\dot{x}$). Depending on the used \textit{Metric Type} to measure $x$, the weight $w$ is calculated next and multiplied by each normalised value ($\dot{x}$). Finally, each weighted normalised value ($\dot{x} * w$) can be aggregated using the \textit{Onion Layer Model} to calculate the \textit{Total Costs} (sum of \textit{Functional Costs} and \textit{Security Costs}) of the measured interaction. The following Figure shows the described SCMF approach for normalising, weighing and aggregating the \textit{Total Costs}:
\vspace{-10pt}
\begin{figure}[H]
	\centering
	\includegraphics[width=\textwidth]{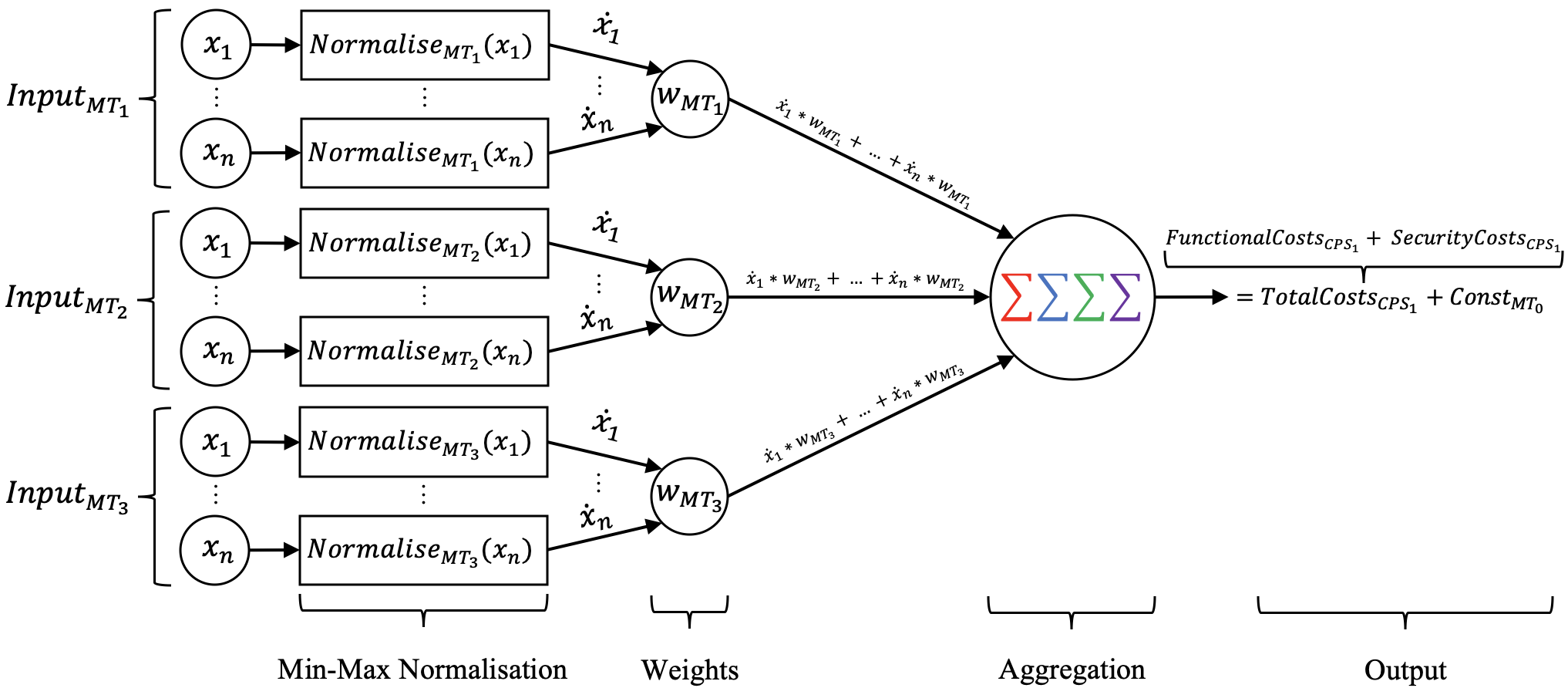}
	\caption{The SCMF for Normalising, Weighting and Aggregating Functional Costs and Security Costs.}
	\Description{Fig3.}
	\label{fig:Fig3}
\end{figure}

\vspace{-10pt}
The following algorithm is based on the approach from Fig. \ref{fig:Fig3} and uses captured task measurement values to normalise, weight  and aggregate the resulting \textit{Functional Costs} and \textit{Security Costs}:

\begin{algorithm}
	\DontPrintSemicolon
	\KwIn{A sequence of measurement objects $X = \langle x_1, x_2, \ldots, x_n \rangle$}
	\KwOut{The normalised and aggregated \textit{Functional} and \textit{Security Costs} of the given input}
	$FunctionalCosts \gets 0, SecurityCosts \gets 0$\;
	\ForEach{$x \in X$}{
		$\dot{x} \gets normalise(x)$\;
		$w \gets calculateWeight(x.getMetricType())$\;
		\eIf{$x.getTask() \in SecurityRelatedTasks$}{$SecurityCosts \gets SecurityCosts + (\dot{x} * w)$\;}{$FunctionalCosts \gets FunctionalCosts + (\dot{x} * w)$\;}
	}
	\Return{FunctionalCosts, SecurityCosts}\;
	\caption{{\sc Cost Aggregation}}
	\label{algo:SC}
\end{algorithm}

\section{Architecture}\label{section:architecture}

In this section, we describe an overall use case that has the main purpose to control the temperature of a physical room. Next, we present the Arrowhead Framework in combination with the Pinpoint APM tool and explain how the overall use case could be set up and measured. In this regard, we first describe the basic idea of the Arrowhead Framework including the functionality and interoperability of its three mandatory core systems. Furthermore, we explain how it can be used for setting up the use case, for drawing the system boundaries of the newly deployed CPS and controlling the behaviour within it. Finally, we show how the performance of the deployed CPS can be measured at runtime by using the Pinpoint APM tool and the PiLogger One. 

\subsection{Overall Use Case}

In many respects, the control system view in Fig. \ref{fig:Fig4} corresponds to one of the most fundamental definitions of CPS. Even though Helen Gill first introduced the term CPS \cite{ref45}, its roots can be traced back to Norbert Wieners definition of the term \textit{cybernetics} \cite{ref42}. He described it as the conjunction of computation and communication, where the control of physical processes is performed in closed-loop feedback. The CPS in Fig. \ref{fig:Fig4} follows a similar principle of two communicating components ($C_1$ and $C_2$) that measure and change their physical environment (if necessary). Furthermore, this interdependency of measuring and changing the physical world duplicates Wieners closed-loop feedback definition of \textit{cybernetics} \cite{ref42}. For these two reasons, we defined an overall use case, where two components interact with each other in a closed-loop. In this regard, the \textit{Closed-Loop Temperature Control}-use case is defined by two components which together are capable of measuring and changing the temperature of a physical room. To do so, one component ($C_1$) controls an air-conditioning system, while the other component ($C_2$) uses a temperature sensor. The following figure shows the overall use case and illustrates the interaction between $C_1$ and $C_2$: 
\vspace{-13pt}
\begin{figure}[H]
	\centering
	\includegraphics[width=\textwidth]{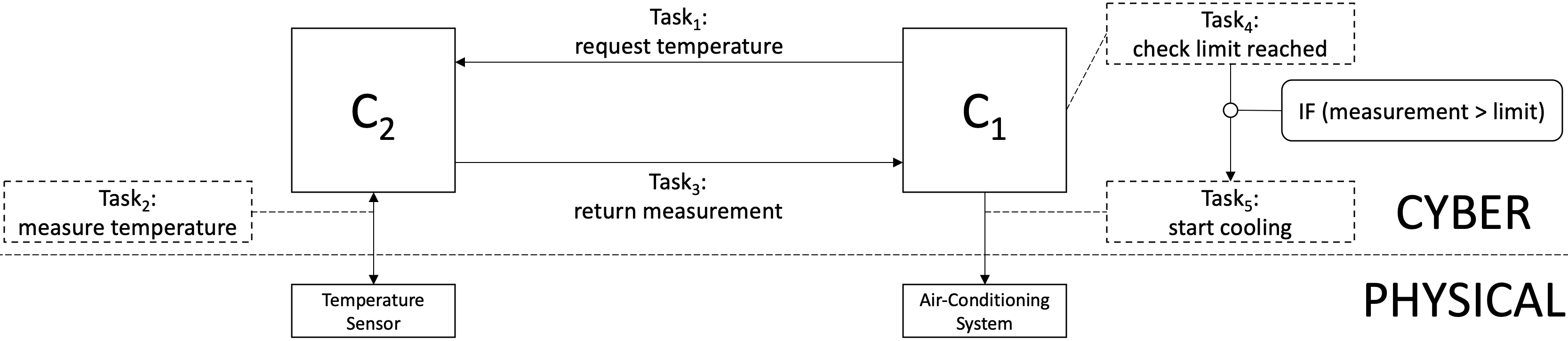}
	\caption{Overall Use Case: Closed-Loop Temperature Control.}
	\Description{CLTC.}
	\label{fig:Fig4}
\end{figure}

\vspace{-5pt}
The interaction between these two components could be described by the following five tasks:
\begin{itemize}
	\item $Task_1$: before $C_1$ decides, to turn on the air-conditioning system, it first requests the room temperature from $C_2$
	\item $Task_2$: $C_2$ uses its sensor to measure the current room temperature
	\item $Task_3$: $C_2$ returns the requested measurement data to $C_1$
	\item $Task_4$: once $C_1$ receives the room temperature it verifies whether a predefined limit has been reached. For instance, if the current temperature is higher than predefined limit the air-conditioning system has to be turned on
	\item $Task_5$: if the predefined limit has been reached, $C_1$ activates the air-conditioning system
\end{itemize}

\vspace{5pt}
This overall use case describes in five basic steps, how a CPS could be set up to measure and change the temperature of a physical room (if necessary). Furthermore, it illustrates the interdependency, where a temperature measurement triggers a change of the physical world which in return again influences the next measurement, and so on. 

\subsection{Arrowhead Framework}
As initially mentioned, two components that are capable of sensing and affecting the physical world must not automatically be called a CPS. The general approach of defining which components are part of/or belong to a specific CPS is to use an IoT-Framework. In general, these frameworks provide some kind of procedure for \textit{enrolling} new components which ensures that they enter an existing CPS in a structured way without compromising it. More precisely, before a new component becomes part of an existing CPS and is allowed to interact with other components, it first has to pass a series of predefined checks. Depending on the IoT-Framework and/or the CPS requirements, the \textit{enrolment} procedure often involves security and authorisation checks, like the verification of a pre-shared key or the validation of exchanged certificates. Some frameworks go even further by verifying certain aspects of the component’s software and hardware features. To give an example, a software check might include verifying which operating system (OS) is running on the component and if the latest patches have been installed. A hardware check could involve verifying, whether a certain chipset is being used which has known security issues. All in all, IoT-Frameworks can be used to define the system boundaries of registered components of a CPS without compromising it.%In summary, IoT-Frameworks can be used to define the system boundaries of all registered components within a specific CPS without compromising its security.

Once the \textit{enrolment} procedure has been successfully completed and the new component becomes part of the CPS, it can start interacting with other (already registered) components. In order to do so, it needs to know which other components are currently part of the same CPS and what they are capable of doing. In addition to that the new component needs to know how to reach (or communicate with) the other existing components. For instance, after the registration the two components from Fig. \ref{fig:Fig4} are part of the same CPS, are there for different reasons and provide different functionalities. One component ($C_1$) is capable of cooling down a physical room, while the other component ($C_2$) provides the functionality of measuring the room’s temperature. However, $C_1$ has no knowledge about the existence of $C_2$ (or any other component within the same CPS) and about its temperature measurement capabilities. Moreover, even if $C_1$ had knowledge about the existence of $C_2$ and which functionality it provides, it still needs to know its communication endpoint or how it can be reached. To ensure that two components can find each other before they start to interact, the IoT-Framework has to fulfil the following additional requirements \cite{ref60}:
\begin{itemize}
	\item managing how a specific component searches for other components within the same CPS, 
	\item enabling data exchange between coupled components, and
	\item regulating which components are allowed to interact with each other.
\end{itemize}

The overall goal of the Arrowhead Framework is to efficiently support the development, deployment and operation of interconnected, cooperative systems using a SOA \cite{ref58} philosophy. In addition to that the Arrowhead Framework addresses the above challenges and requirements by providing an \textit{On-Boarding Procedure} \cite{ref46} and three mandatory core systems \cite{ref60}. Depending on the configuration of the CPS (pre-shared key or certificate), the \textit{On-Boarding Procedure} involves different verification, authentication and authorisation steps, before allowing a new component into an existing CPS. By splitting up the procedure in different steps involving different checks a so called \textit{Chain of Trust} is formed \cite{ref46}. This ensures that only a component that completed all checks of this \textit{chain} is allowed to enter the CPS and interact with other components within it. Thus, if only one check returns an undesired result, the entire \textit{Chain of Trust} is considered as invalid and the component is prohibited from entering the CPS.

Besides managing how new components are \textit{on-boarded}, the Arrowhead Framework uses three mandatory core systems to form a so-called \textit{local cloud}. These core systems are the \textit{Service Registry System}, the \textit{Authorisation System} and the \textit{Orchestrator System} which together form the minimum required system (mandatory) to deploy an \textit{Arrowhead compliant} CPS. According to \citeauthor{ref36} (\citeyear{ref36}), the \textit{local cloud} concept is based on the idea that geographically local automation tasks should be encapsulated and protected. Moreover, deploying the three mandatory core systems automatically creates a \textit{local cloud} that includes all nearby components (drawing a system boundary) and supports them in performing the desired automation tasks. Furthermore, this \textit{local cloud} provides security controls (e.g.: encrypted communication, authorisation, authentication, etc.) to protect the components from outside influences \cite{ref36}. In addition to that, each \textit{local cloud} deployment is capable of interconnecting with other \textit{local cloud} deployments to create large-scale System of Systems (SoS) \cite{ref59}. As a result, the Arrowhead Framework is capable of managing both small groups of \textit{local} components, as well as large-scaled interconnected meshes of CPS.

To successfully manage IoT-components and enable task automation within a \textit{local cloud}, the Arrowhead Framework uses a SOA and reduces everything within a CPS to a service. Based on that idea the CPS in Fig. \ref{fig:Fig4} would reduce $C_1$ to the \textit{Air-Conditioning Service} and $C_2$ to the \textit{Temperature Measurement Service}. More precisely, in the context of the Arrowhead Framework a service is what is used to exchange information from providing to consuming system. For instance, as shown in Fig. \ref{fig:Fig4}, $C_1$ is consuming the \textit{Temperature Measurement Service} which is provided by $C_2$. At the same time $C_1$ is also providing the \textit{Air-Conditioning Service} which depends on the measurement data from \textit{Temperature Measurement Service} provided by $C_2$. Generally speaking, the idea of the Arrowhead Framework is to deploy a \textit{local cloud} to define the system boundary of nearby (local) IoT-components. In addition to that each \textit{on-boarded} component can provide one or more services to the \textit{local cloud} which can be consumed by other components. At the same time the same component can also consume any service provided by any other component within the same \textit{local cloud} deployment. Summarizing, within a \textit{local cloud} each component is defined by its provided services which can be consumed by any other component within the same CPS.

The mandatory core systems play a key role for managing services of a CPS and controlling how interactions are executed between two or more components. In this regard, each core system is designed in a way that its provided services are interdependent. Furthermore, these services can be consumed by the other core systems as well as any other component within the same CPS. As a result, the mandatory core systems provide basic functionalities that support the automation of tasks to successfully execute interactions between two or more components. The following sections describe the idea, basic principles and responsibilities of the three mandatory core systems \cite{ref60}:

\subsubsection{Service Registry System}
This system provides services for registering new services and looking up currently active offered services (\textit{Service Discovery Service}). In order to register its own services to the \textit{local cloud}, a newly introduced component \textit{publishes} them in the last step of the \textit{On-Boarding Procedure}. This registration step uses the \textit{Service Registry} to store all relevant information about a service (e.g. metadata, network address, port, path, etc.) into a database. The \textit{Service Discovery Service} can later be used to search for (or lookup) existing services within the same database. However, it is worth mentioning that instead of directly consuming the \textit{Service Discovery Service} the standard procedure for looking up a service is to send a request to the \textit{Orchestrator System}. Since the \textit{Orchestration Process} involves interacting with the other two mandatory core systems, it takes care of locating all suitable services and performing an authorisation check.

\subsubsection{Authorisation System}
This system provides services for creating authorisation rules that describe which components are allowed to consume which services within a specific CPS (intra-cloud access rule). Furthermore, it is also possible to create authorisation rules that describe which other \textit{local clouds} are allowed to consume which services from a specific CPS. Therefore, by providing an \textit{Authorisation Control Service}, this systems purpose is to verify which component is authorised to consume which service within its own \textit{local cloud} and other interconnected CPS. Similar to the \textit{Service Discovery Service}, the \textit{Authorisation Control Service} is used during the \textit{Orchestration Process} to verify, whether a component is allowed to consume the requested service.

\subsubsection{Orchestrator System}
The primary purpose of this system is to provide the communication endpoints of other services within the \textit{local cloud} including all necessary information to enable interactions between components. Thus, the \textit{Orchestrator System} represents the single point of contact for processing any kind of requests sent by any component within the CPS. More precisely, each request sent to the \textit{Orchestrator System} will first be analysed and then forwarded to the right systems within the \textit{local cloud} (if necessary). In other words, this system handles all requests internally and returns the results (or guiding actions) to the requesting component. For instance, $C_1$ would send a request to the \textit{Orchestrator System} asking for a component with a \textit{Temperature Measurement Service}. This request is first forwarded to the \textit{Service Registry System} which returns all required information about $C_2$ and its service. Before returning this information to $C_1$, the \textit{Orchestrator System} then asks the \textit{Authorisation System}, if $C_1$ is allowed to consume the service provided by $C_2$. After the authorisation check is completed successfully (meaning: $C_1$ is allowed to interact with $C_2$), all relevant metadata and endpoint information of the requested\textit{Temperature Measurement Service} is sent to $C_1$. Finally, $C_1$ receives all necessary information to start interacting with $C_2$ by consuming the requested \textit{ Temperature Measurement Service}. Summarizing, the \textit{Orchestration Process} forms a link between the mandatory core systems and performs task automation of \textit{discovering requested services} and \textit{performing authorisation checks} before returning the requested information.

\subsection{Pinpoint}

Pinpoint is an open source APM tool for measuring the performance of large-scale distributed systems using an agent-based approach. Inspired by Dapper \cite{ref52}, Pinpoint provides a solution to help analyse the overall structure of a system and how its components within are interconnected by tracing transactions across distributed applications. More precisely, each component of a system can be instrumented by using an agent that monitors the running application at runtime and provides code-level information about any executed transaction \cite{ref47}. Furthermore, according to \cite{ref53} the agent-based approach has a minimal impact on the performance of the underlying system by an approximate 3\% increase in resource usage. Summarizing, Pinpoint is a suitable tool for measuring the performance of an Arrowhead managed CPS for the following reasons:
\begin{itemize}
	\item the Arrowhead Framework including its mandatory core systems and Pinpoint including its agent are both written in Java which makes them compatible
	\item the agent-based approach is capable of instrumenting all components of a \textit{local cloud} without changing a single line of code
	\item Pinpoint uses a set of different metrics to measure the performance of executed tasks
	\item using Pinpoint for runtime measurements has an insignificant impact on the performance
\end{itemize}

\vspace{9pt}
As previously mentioned, Pinpoint is modelled after Google’s Dapper and uses a combination of \textit{Transaction Tracing} and \textit{Bytecode Instrumentation} techniques for measuring the performance. The purpose of tracing transactions is to analyse relationships between two nodes in a distributed system and identify when a message is sent from one node to the other. In this regard, the Pinpoint tracing technique modified the Dapper approach by adding an application-level tag in the call header in order to trace distributed transactions at a remote call. Besides that, Pinpoint adopted the bytecode instrumentation technique to enable performance measurements at runtime without having to change a single line of code. Instead, an agent is used to intervene Remote Procedure Calls (RPCs) and to handle tag information automatically. Another advantage of automatic tracing is that more precise data can be collected to more information in the bytecode. After the measurement the agent sends the trace data to the \textit{Pinpoint Collector} to store it in a database. From there, all measurements can be extracted, the raw data normalised, and the \textit{Total Costs} calculated. The following figure demonstrates the basic principle of transaction tracing 
using an automatic agent-based approach:
\vspace{-5pt}
\begin{figure}[H]
	\centering
	\includegraphics[width=\textwidth]{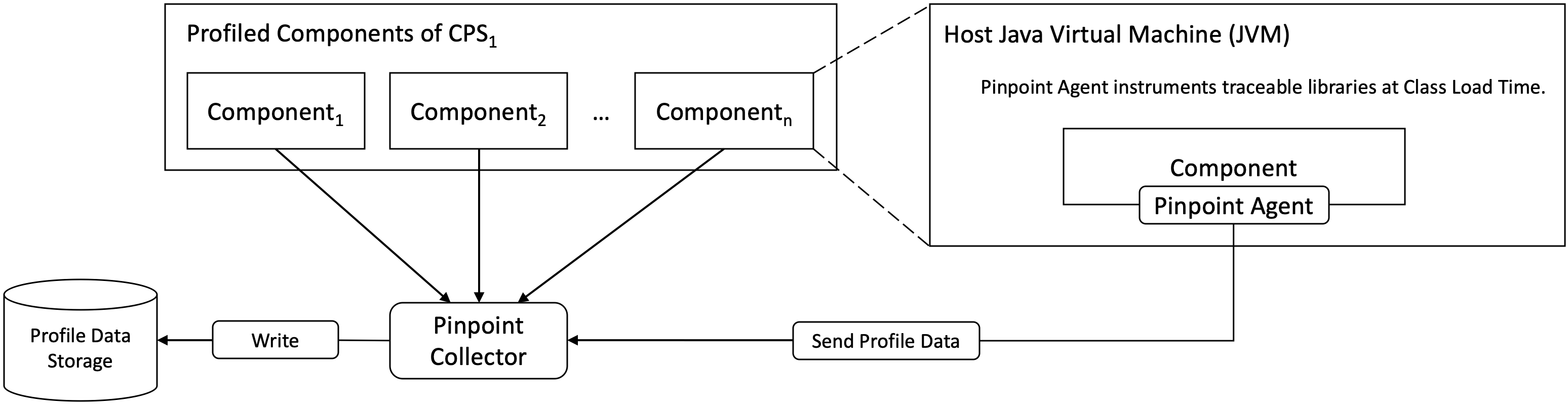}
	\caption{Pinpoint Architecture for Measuring the Performance of Components (adapted from \citeauthor{ref47}, 2020).}
	\Description{CLTC-AH.}
	\label{fig:Fig5}
\end{figure}

\subsection{Experimental Testbed}

Based on the overall use case from Fig. \ref{fig:Fig4}, we setup an experimental testbed by combining the Arrowhead Framework, the Pinpoint APM tool and the PiLogger One board. In this regard, we deployed the mandatory core systems to create a \textit{local cloud} and used the \textit{On-Boarding Procedure} to register the components $C_1$ and $C_2$. Furthermore, all necessary authorisation rules were created to allow interactions between the three mandatory core systems and the two components. Moreover, to be able to effectively compare measurement results of different components and have full control of each test run the testbed was configured as follows:
\begin{itemize}
	\item each of the three core systems, $C_1$ and $C_2$ were running on a separate \textit{Raspberry Pi 3 Model B+}
	\item each component was connected over a Local Area Network (LAN) via Ethernet cable
	\item each Raspberry Pi used a minimal image based on Debian Buster (\textit{Raspbian Buster Lite}) as OS
	\item wireless LAN, Bluetooth, the High Definition Multimedia Interface (HDMI) and the Light Emitting Diodes (LEDs) were disabled
\end{itemize}

\vspace{5pt}
Since each component was running on a Raspberry Pi the comparison of the measurement results is easier, due to the same hardware specifications. Another advantage is that interconnecting all devices over a LAN via Ethernet cable gives more control over the network and reduces potential disturbances from outside. Besides that, to conserve existing computing resources on each Raspberry Pi as much as possible Raspbian Buster Lite was installed and used as OS. Finally, all unused interfaces (HDMI, Bluetooth and LEDs) have been disabled to further reduce any loss in performance and improve the power consumption of each component. 

As shown in Fig. \ref{fig:Fig6}, we used one representative metric from each metric type to measure the \textit{Functional Costs} and the \textit{Security Costs} of a deployed CPS. In this regard, the execution time of tasks ($Metric Type_1$) and the CPU-usage ($Metric Type_2$) were measured using a Pinpoint Agent for each component. Each agent was used to instrument the bytecode of the control logic of each component and measure the performance of each task. After that the measurement raw data is sent to the Pinpoint Collector which stores it into a database (Performance Raw Data). 

Besides measuring the execution time ($Metric Type_1$) and CPU-usage ($Metric Type_2$), the power consumption ($Metric Type_3$) was measured by adding a PiLogger One \cite{ref61} to each component. By doing so the PiLogger One adds an autonomous working four-channel measuring system connected via Inter-Integrated Circuit ($I^2C$) bus to the Raspberry Pi. In addition to that the PiLogger One comes with its own timebase which allows precise timed intervals between measurements independent of the actual activity running on the Raspberry Pi. The concept of an autonomous peripheral unit enables measuring the power consumption (voltage and current) of components using a Raspberry Pi. Similar to the Pinpoint Collector, the measured power consumption data can be retrieved and stored in a separate database (Power Consumption Raw Data). 

In the end, the measured performance and power consumption raw data can be extracted from both databases to calculate (normalise and aggregate) the \textit{Functional Costs} and the \textit{Security Costs} using the \textit{Onion Layer Model}. Summarizing, an experimental testbed consisting of the \textit{Arrowhead Framework}, the \textit{Pinpoint APM} tool and the \textit{PiLogger One} extension card enables configuring CPS, running interactions and measuring their performance. The following figure shows the testbed setup including all components (mandatory core systems, $C_1$ and $C_2$), Pinpoint Agents and PiLogger One boards:
\vspace{-5pt}
\begin{figure}[H]
	\centering
	\includegraphics[width=13cm]{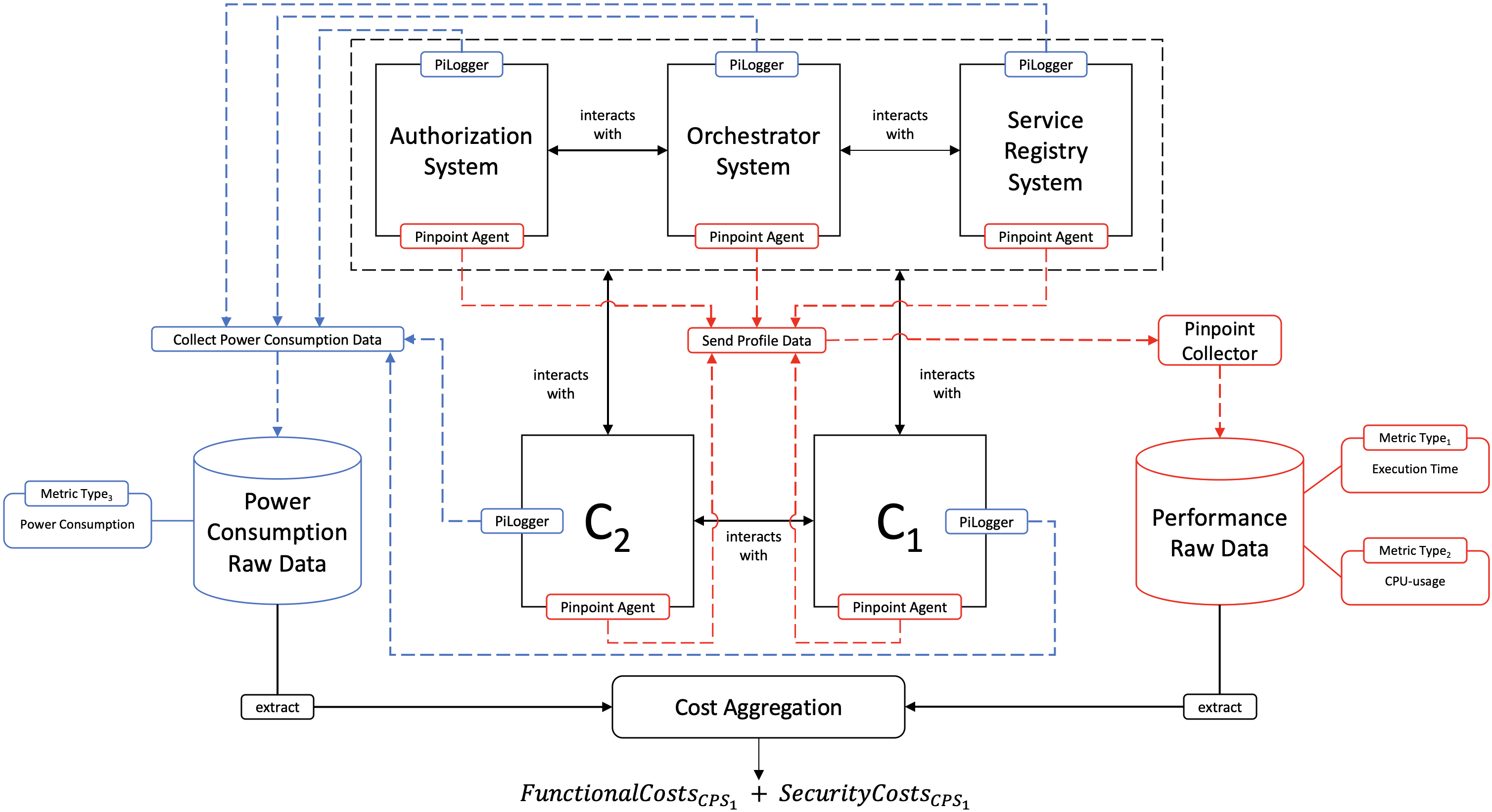}
	\caption{Setup of the Experimental Testbed using the Arrowhead Framework, Pinpoint APM and PiLogger One.}
	\Description{Testbed}
	\label{fig:Fig6}
\end{figure}

\section{Experimental Study}\label{section:experiment}

In this section, we show how the overall use case from Fig. \ref{fig:Fig4} can be measured using the Experimental Testbed setup as shown in Fig. \ref{fig:Fig6}. In this regard, we present two possible solutions (\textit{Use Case 1}, \textit{Use Case 2}) for implementing the \textit{Closed-Loop Temperature Control} interaction of the overall use case using the Arrowhead Framework. Next, we define four different workloads ($WL_{1.1}$, $WL_{1.2}$, $WL_{2.1}$, $WL_{2.2}$) and explain how they have been used to measure, normalise and aggregate the \textit{Functional Costs} and the \textit{Security Costs} during the experimental study.

\subsection{Experimental Settings}

As shown in Fig. \ref{fig:Fig4}, the overall goal of the \textit{Closed-Loop Temperature Control} interaction is to measure, analyse and change the temperature of a physical room (if necessary). To achieve this goal, the interaction involves one component that controls an air-conditioning system ($C_1$) and another one that uses a temperature sensor ($C_2$). To successfully implement this interaction, we first deployed an Arrowhead \textit{local cloud} with its three mandatory core systems. Next, we used the \textit{On-Boarding Procedure} to register the \textit{Air-Conditioning Service} of $C_1$ and the \textit{Temperature Measurement Service} of $C_2$ to the database of the \textit{Service Registry System}. Additionally, we created a rule that defines that $C_1$ and $C_2$ are authorised to interact with each other. At this point, $C_1$ and $C_2$ belong to the same CPS but have no information about each other. 

According to the Arrowhead Framework a component within a \textit{local cloud} can either be a \textit{Producer} or a \textit{Consumer}. As shown in Fig. \ref{fig:Fig4}, $C_2$ is a \textit{Producer} that offers a \textit{Temperature Measurement Service} that can be consumed by other components. On the contrary, $C_1$ is a \textit{Consumer} that consumes the services provided by $C_2$ to decide whether the air-conditioning system needs to be activated or not. As described by the Arrowhead Framework, the recommended practice for designing interactions is that the \textit{Consumer} has to \textit{lookup} the \textit{Producer} by sending a request to the \textit{Orchestrator System}. However, it would not be wrong to start the interaction the other way around either, where the \textit{Producer} ($C_2$) sends a request to the \textit{Orchestrator System} looking for the \textit{Consumer}. Thus, to successfully execute the \textit{Closed-Loop Temperature Control} interaction, we implemented both use cases and measured their performance to identify which use case is more efficient.

\subsubsection{Use Case 1 - Consumer to Producer}

As already mentioned, after $C_1$ and $C_2$ have been successfully \textit{on-boarded} they cannot start to interact with each other right away. Instead, they can look for the other component by sending a request to the \textit{Orchestrator System} which takes care of the \textit{lookup} and \textit{authorisation check} within the \textit{local cloud}. In this regard, the \textit{Orchestrator System} first forwards the original request to the \textit{Service Registry System} which searches its database to \textit{lookup} the requested component ($C_2$) with the temperature sensor. If there is such a component, the \textit{Orchestrator System} asks the \textit{Authorisation System}, if there is a rule that allows the requesting component ($C_1$) and  the requested component ($C_2$) to interact with each other. Upon a successful authorisation check, the \textit{Orchestrator System} returns the endpoint data of $C_2$ to $C_1$ which now triggers the interaction between those two components. According to the best-practice guidelines provided by the Arrowhead Framework, $C_1$ has to initiate the \textit{Closed-Loop Temperature Control} interaction, by requesting the endpoint of $C_2$ from the \textit{Orchestrator System}. Fig. \ref{fig:Fig7} shows the steps of the \textit{Closed-Loop Temperature Control} interaction following the \textit{Consumer to Producer} approach:
\vspace{-5pt}
\begin{figure}[H]
	\centering
	\includegraphics[width=\textwidth]{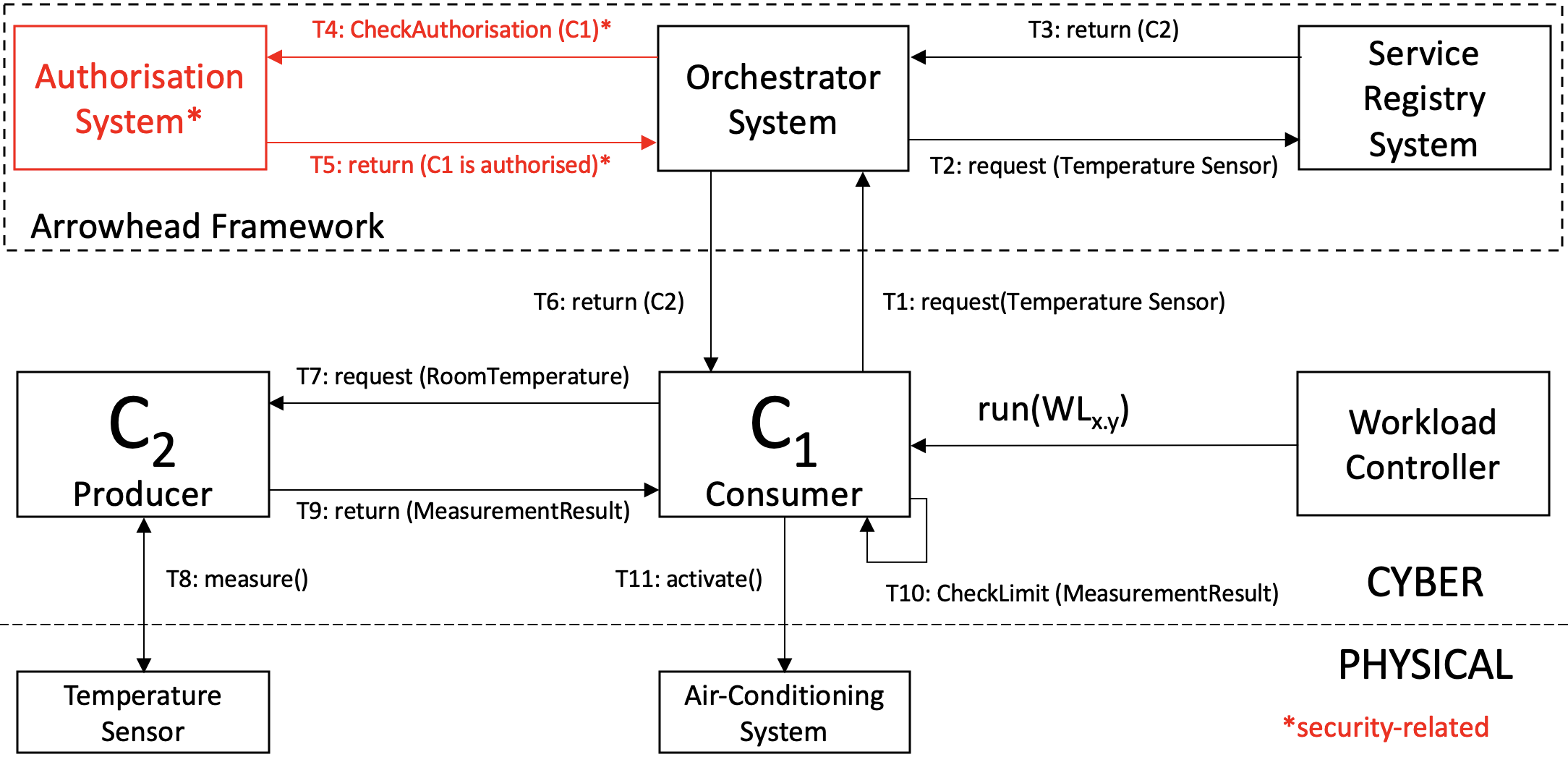}
	\caption{\textit{Closed-Loop Temperature Control} Interaction initiated by $C_1$ (\textit{Use Case 1 - Consumer to Producer}).}
	\label{fig:Fig7}
\end{figure}

\vspace{-5pt}
As shown in Fig. \ref{eq:7}, each component performs different tasks to reach the overall goal of measuring and controlling the temperature of a physical room. In this regard, the tasks performed by $C_1$, $C_2$ and the \textit{Service Registry System} are mainly \textit{functional} tasks, while the \textit{Authorisation System} is mainly responsible for \textit{security-related} tasks. The \textit{Orchestrator System} is a hybrid system, since it performs \textit{functional} tasks when interacting with the \textit{Service Registry System}, but also performs \textit{security-related} tasks when communicating with the \textit{Authorisation System}. To measure the \textit{Security Costs} of \textit{Use Case 1}, all tasks performed by the \textit{Authorisation System} and its interaction with the \textit{Orchestrator System} have to be included in the SCMF. The following listing describes the purpose and functionality of each task from Fig. \ref{fig:Fig7} (\textcolor{xred}{red = security-related tasks}):
\begin{itemize}
	\item$T_1$: $C_1$ requests a component with a temperature sensor from the \textit{Orchestrator System}
	\item$T_2$: the \textit{Orchestrator System} forwards the request to the \textit{Service Registry System}
	\item$T_3$: the \textit{Service Registry System} searches its database and returns $C_2$
	\textcolor{xred}{\item$T_4$: the \textit{Orchestrator System} asks the \textit{Authorisation System} if $C_1$ is allowed to consume $C_2$}
	\textcolor{xred}{\item$T_5$: the \textit{Authorisation System} searches all authorisation rules and returns \textit{$C_1$ is authorised}}
	\item$T_6$: the \textit{Orchestrator System} returns the endpoint of $C_2$ to $C_1$
	\item$T_7$: $C_1$ requests the temperature of the room from $C_2$
	\item$T_8$: $C_2$ uses its sensor to measure the temperature
	\item$T_9$: $C_2$ returns the measured value to $C_1$
	\item$T_{10}$: $C_1$ checks if the measured value is greater than a predefined limit (limit = 25\textdegree{}C)
	\item$T_{11}$: if the measurement is greater than the limit $C_1$ activates the air-conditioning system
\end{itemize}

\subsubsection{Use Case 2 - Producer to Consumer}
Even though the use case in Fig. \ref{fig:Fig7} is the recommended way it has one crucial disadvantage. The entire interaction is executed from $T_1$ to $T_{10}$ even if the measured room temperature is below the predefined limit of 25\textdegree{}C. This means that when the measured room temperature is below 25\textdegree{}C, these ten tasks have been executed without causing any effect. To overcome this issue, the second use case is designed in a way that $C_2$ (instead of $C_1$) initiates the entire interaction. Fig. 8 shows the steps of the \textit{Closed-Loop Temperature Control} interaction following the redesigned \textit{Producer to Consumer} approach:
\vspace{-5pt}
\begin{figure}[H]
	\centering
	\includegraphics[width=\textwidth]{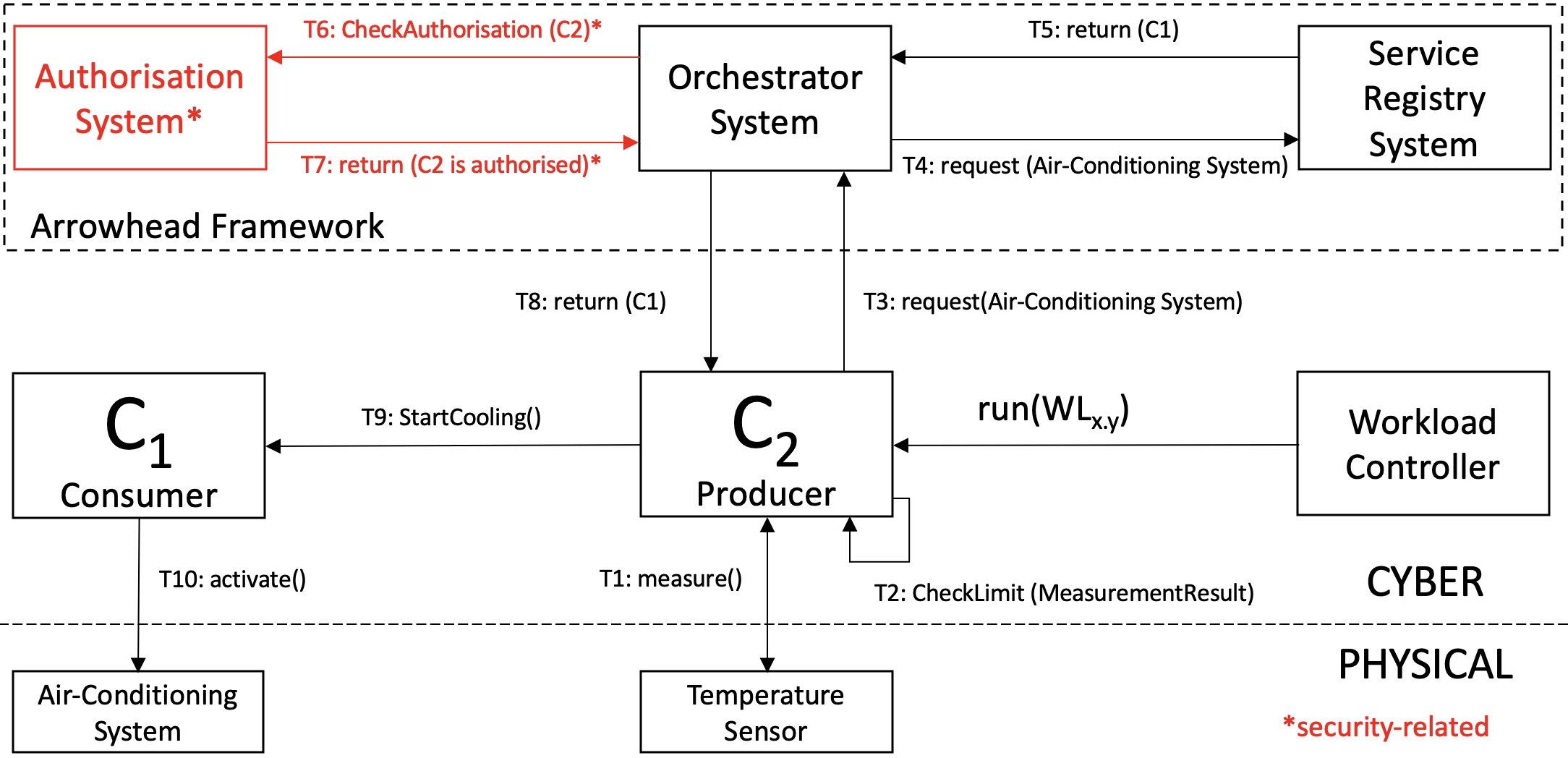}
	\caption{\textit{Closed-Loop Temperature Control} Interaction initiated by $C_2$ (\textit{Use Case 2 - Producer to Consumer}).}
	\label{fig:Fig8}
\end{figure}

\vspace{-5pt}
As shown in Fig. \ref{fig:Fig8}, the first two tasks of $C_2$ are to measure the room temperature and check, whether the predefined limit has been reached, or not. These two steps are executed in a continuous loop, until finally the measured temperature is higher than 25\textdegree{}C. Only then all the other tasks from the \textit{Closed-Loop Temperature Control} interaction are executed. The advantage of this approach is that the additional tasks from $T_3$ to $T_{10}$ are only executed, when they are really required. Besides changing the initiating component (from $C_1$ to $C_2$) all other tasks are executed the same way as before. The only difference is that the \textit{Service Registry System} returns the endpoint of $C_1$ (instead of $C_2$) and the \textit{Authorisation System} checks the authorisation of $C_2$ (instead of $C_1$). This also includes the components responsible for executing the \textit{security-related} tasks (\textit{Authorisation System} and its interaction with the \textit{Orchestrator System}). The following listing describes the purpose and functionality of each task from Fig. \ref{fig:Fig8} (\textcolor{xred}{red = security-related tasks}):
\begin{itemize}
	\item$T_1$: $C_2$ uses its sensor to measure the temperature
	\item$T_2$: $C_2$ checks if the measured value is greater than a predefined limit (limit = 25\textdegree{}C)
	\item$T_3$: if the measurement is greater than the limit, $C_2$ requests a component with an air-conditioning system from the \textit{Orchestrator System}
	\item$T_4$: the \textit{Orchestrator System} forwards the request to the \textit{Service Registry System}
	\item$T_5$: the \textit{Service Registry System} searches its database and returns $C_1$
	\textcolor{xred}{\item$T_6$: the \textit{Orchestrator System} asks the \textit{Authorisation System} if $C_2$ is allowed to consume $C_1$}
	\textcolor{xred}{\item$T_7$: the \textit{Authorisation System} searches all authorisation rules and returns \textit{$C_2$ is authorised}}
	\item$T_8$: the \textit{Orchestrator System} returns the endpoint of $C_1$ to $C_2$
	\item$T_9$: $C_2$ sends a requests to $C_1$ to start cooling down the room
	\item$T_{10}$: $C_1$ activates the air-conditioning system
\end{itemize}

\subsubsection{Workload Controller}

As described above, both use cases use different approaches to achieve the overall goal of the \textit{Closed-Loop Temperature Control} interaction. The first use case follows the recommended \textit{Consumer to Producer} approach, while the second use case initiates the interaction the other way around (\textit{Producer to Consumer}). To measure the \textit{Functional Costs} and the \textit{Security Costs} of both use cases, we designed a controlled experimental study where a \textit{Workload Controller} started each use case by using the following four workloads: 
\begin{table}[H]
	\caption{Workloads used by the Workload Controller for the Experimental Study}
	\label{tab:workloads}
	\begin{tabular}{ccccc}
		\toprule
		WL&Use Case&Runs&Temperature&Protocol\\
		\midrule
		\multirow{2}{*}{$WL_{1.1}$}&\multirow{2}{*}{Use Case 1}&25&Measurement \textless 25\textdegree{}C&\multirow{2}{*}{HTTPS}\\
		&&25&Measurement \textgreater 25\textdegree{}C&\\
		\midrule
		\multirow{2}{*}{$WL_{1.2}$}&\multirow{2}{*}{Use Case 1}&25&Measurement \textless 25\textdegree{}C&\multirow{2}{*}{HTTP}\\
		&&25&Measurement \textgreater 25\textdegree{}C&\\
		\midrule
		\multirow{2}{*}{$WL_{2.1}$}&\multirow{2}{*}{Use Case 2}&25&Measurement \textless 25\textdegree{}C&\multirow{2}{*}{HTTPS}\\
		&&25&Measurement \textgreater 25\textdegree{}C&\\
		\midrule
		\multirow{2}{*}{$WL_{2.2}$}&\multirow{2}{*}{Use Case 2}&25&Measurement \textless 25\textdegree{}C&\multirow{2}{*}{HTTP}\\
		&&25&Measurement \textgreater 25\textdegree{}C&\\
		\bottomrule
	\end{tabular}
\end{table}

As shown in Table \ref{tab:workloads}, each workload was executed on either the first, or the second use case in a total of 50 runs per workload. In addition to the first 25 runs generated a value below 25\textdegree{}C every time the temperature sensor of $C_2$ was used to measure the room temperature. The remaining 25 runs of the same workload generated a measurement value above 25\textdegree{}C. The reason for that was to split up the total runs into two equal groups and evaluate the \textit{Functional Costs} and \textit{Security Costs} of both temperature measurement outcomes for both use cases. Besides that, each workload defined that the components either used a secure (HTTPS) or insecure (HTTP) communication protocol when exchanging messages with each other. Similar to \cite{ref50}, we used the secure/insecure protocol option in each workload to measure the costs of the "S" in HTTPS. This allows to quantify the \textit{Security Costs} of the secure communication protocol by comparing the measurement results with its insecure counterpart. In addition to the four workloads in Table \ref{tab:workloads}, we used the following metrics to measure the performance  of each interaction:
\vspace{-5pt}
\begin{table}[H]
	\caption{Performance Metrics for Measuring \textit{Functional Costs} and \textit{Security Costs} for each Workload}
	\label{tab:metrics}
	\begin{tabular}{ccc}
		\toprule
		Metric Type&Metric&Unit\\
		\midrule
		$MT_0$&Algorithmic Time Complexity (Big-O)&-\\
		$MT_1$&Duration of executing a specific task&Milliseconds (ms)\\
		$MT_2$&CPU-usage&Percent (\%)\\
		$MT_3$&Power consumption&Milliwattseconds (mWs)\\
		\bottomrule
	\end{tabular}
\end{table}

\vspace{-5pt}
To show how to successfully measure the \textit{Total Costs} of both use cases by using the SCMF, we defined one representative metric for each of the four \textit{Metric Types}. As shown in Table \ref{tab:metrics}, these metrics were used to measure the performance while executing each of the defined workloads from Table \ref{tab:workloads}. After that each raw data measurement value was normalised, weighted and aggregated following the SCMF approach from Fig. \ref{fig:Fig3}. As a result, the controlled experimental setup enables to use predefined workloads in combination with four different metrics to evaluate which use case executes the \textit{Closed-Loop Temperature Control} interaction more efficiently.

\subsection{Total Cost Evaluation and Security Cost Extraction of Use Case 1 and Use Case 2}\label{subsection:SCMF}

Following the presented SCMF approach as shown in Fig. \ref{fig:Fig3}, the first step is to measure and calculate the raw data measurement results. In this regard we used the first metric from Table \ref{tab:metrics} to estimate the algorithmic time complexity of each component using the Big-O notation. As previously described, due to the \textit{Halting Problem} this step has to be done manually by analysing the program code of the Arrowhead Framework for each of the five components. The following table shows the results of the algorithmic time complexity estimation for each workload:
\vspace{-5pt}
\begin{table}[H]
	\caption{Algorithmic Time Complexity Estimation per Workload}
	\label{tab:atce}
	\begin{tabular}{ccccc}
		\toprule
		\multirow{2}{*}{Component}&\multicolumn{4}{c}{Workloads}\\
		&$WL_{1.1}$&$WL_{1.2}$&$WL_{2.1}$&$WL_{2.2}$\\
		\midrule
		C$_1$&n&n&1&1\\
		C$_2$&1&1&n&n\\
		Orchestrator System&n&n&n&n\\
		Service Registry System&n&n&n&n\\
		Authorisation System&n&n&n&n\\
		\bottomrule
	\end{tabular}
\end{table}

\vspace{-5pt}
As shown in Table \ref{tab:atce}, the algorithmic time complexity of the three mandatory core systems (\textit{Orchestrator System}, \textit{Service Registry System}, \textit{Authorisation System}) is $f_{MT_0}(n) = n$ for every workload. This is an expected result, since the three core systems executed the same tasks in the same order using the same algorithms in both use cases. The difference in algorithmic time complexity can be shown when looking at $C_1$ and $C_2$. Depending on the use case and which component is the initiator of the entire \textit{Closed-Loop Temperature Control} interaction, the algorithm of the initiating component is shown to be more complex. For instance, in \textit{Use Case 1} where $C_1$ starts the interaction by sending a request to the \textit{Orchestrator System}, its algorithmic time complexity is $f_{MT_0}(n) = n$. For the same use case $C_2$ has an algorithmic time complexity of $f_{MT_0}(n) = 1$. This result is reversed in Use Case 2, since the initiating component changes form $C_1$ to $C_2$.

After classifying the algorithmic time complexity, the next step is to measure and calculate the raw data results at runtime for the \textit{Metric Types} $MT_1$, $MT_2$, $MT_3$. As shown in Fig. \ref{fig:Fig6}, the Pinpoint Agents are used to measure the duration ($MT_1$) and CPU-usage ($MT_2$) for each task of each component. In addition to that the PiLogger One board is attached to each Raspberry Pi to measure the total power consumption ($MT_3$) of all components. This means that the equations (\ref{eq:3}) and (\ref{eq:4}) are integrated in the Pinpoint Agent to calculate the raw data measurement results for the duration and CPU-usage at runtime. To show in an example how these two equations are used in practice, let’s assume that component $c$ performed two \textit{security-related} tasks $s_1$ and $s_2$. The following example shows, how to measure the duration of these two tasks:
\vspace{-5pt}
\begin{table}[H]
	\centering
	\begin{tabular}{cr}
		$X_{1MT_1} = 10ms - \textcolor{white}{0}0ms = 10ms$\text{, } & \multirow{2}{*}{as described in (\ref{eq:3})}\\
		$X_{2MT_1} = 30ms - 10ms = 20ms$\text{, } & \\
	\end{tabular}
\end{table}

\vspace{-5pt}
As described in (\ref{eq:3}), calculating the duration of a task requires to capture a timestamp right before and one right after the execution of the task. The difference of those two capture points (start and end) is the calculated raw measurement data for $MT_1$. On the contrary, measuring the CPU-usage ($MT_2$) requires many sample snapshots, while a certain task has been executed. The number of these snapshots can differ depending on how long the execution time of the task was, or in other words how much time there was to create snapshots. Thus, let's again assume that the number of snapshots created during the execution of $s_1$ was $m_1 = 3$, while the number of snapshots for $s_2$ was $m_2 = 5$. The following example shows, how to measure the CPU-usage of $s_1$ and $s_2$:
\vspace{-5pt}
\begin{table}[H]
	\centering
	\begin{tabular}{cr}
		$X_{1MT_2c} = \frac{(1.5\% + 2.0\% + 1.0\%) + (2.3\% + 2.1\% + 1.8\% + 2.7\% + 1.5\%)}{3 + 5} = (1.825\%)$ & as described in (\ref{eq:4})\\
	\end{tabular}
\end{table}

\vspace{-5pt}
As described in (\ref{eq:4}), to calculate the CPU-usage of a component the overall mean of the of used CPU of each task has to be included. This means that the sum of all sample measurements of all tasks is divided by the sum of the number of created snapshots. By doing so the different sample sizes ($m_1 = 3$ and $m_2 = 5$) of all sample measurement snapshots are combined when calculating the overall mean of all tasks of a component. The resulting overall mean CPU-usage represents a number that expresses how much CPU was utilised by a component on average while performing all its \textit{security-related} tasks. 

Similar to the Pinpoint Agents that use the equations (\ref{eq:3}) and (\ref{eq:4}) to calculate the duration and CPU-usage, the PiLogger One board uses equation (\ref{eq:5}) to calculate the power consumption. In this regard, the integrated software that controls the measurement board attached to each Raspberry Pi calculates the total power consumption while a component is performing all tasks. The following example shows how to measure the total power consumption of component $c$:
\vspace{-5pt}
\begin{table}[H]
	\centering
	\begin{tabular}{cr}
		$X_{1MT_3c} = 0.5mW * 0.01s = 0.005mWs$, & \multirow{2}{*}{as described in (\ref{eq:5})}\\
		$X_{2MT_3c} = 0.5mW * 0.02s = 0.010mWs$, & \\
	\end{tabular}
\end{table}

\vspace{-5pt}
As shown in the example above, the PiLogger One board measured $0.5mW$ while component $c$ performed the two \textit{security-related} tasks $s_1$ and $s_2$. To calculate the total power consumption (e.g. $mWs$) for component $c$ the previously measured execution times ($X_{1MT_1} = 10ms = 0.01s$ and $X_{2MT_1} = 20ms = 0.02s$) have to be multiplied by the $0.5mW$. The same approach, as shown in these three examples, was used to measure and calculate the raw data measurement results for all components while executing the defined workloads. After that the next step of the SCMF is to normalise each raw data value $x$ to calculate the normalised value $\dot{x}$ by using equation (\ref{eq:7}). As mentioned before, the unity-based \textit{MIN-MAX Normalisation} brings all raw data values with different units into the range $[a,b]$. In this article, we defined the lower bound $a = 0$ and the upper bound $b = 1$, to normalise all raw data value between the range of $[0,1]$. In addition to, and as described in (\ref{eq:7}), we define the following lower and upper bounds for each metric type:
\vspace{-5pt}
\begin{table}[H]
	\caption{Lower Bound (MIN) and Upper Bound (MAX) for each Metric Type}
	\label{tab:mtlu}
	\begin{tabular}{ccc}
		\toprule
		Metric Type & $MIN_{MT_j}$ & $MAX_{MT_j}$ \\
		\midrule
		$MT_1$ & 0 & 1000\\
		$MT_2$ & 0 & 100\\
		$MT_3$ & 0 & 10\\
		\bottomrule
	\end{tabular}
\end{table}

\vspace{-5pt}
Generally speaking, the lower and upper bounds have to be assigned a value where the measured raw data value $x$ is within the range of $MIN_{MT_j} \textless x \textless MAX_{MT_j}$. As shown in Table \ref{tab:mtlu}, we defined the lower and upper bounds to represent extreme values that cannot be reached by any measurement result. For instance, the Arrowhead Framework has a constant that sets the connection timeout to $1000ms$. In other words, when communicating with a component they must reply within $1000ms$, otherwise the connection is dropped. For that reason we choose to set the upper bound for $MT_1$ to $1000$. Similar to that we defined the upper bound of $MT_2$ to $100$, since it is virtually impossible to use more than $100\%$ of a CPU. A Raspberry Pi that is running using $100\%$ CPU and running 24 hours a day for an entire year, has a power consumption of approximately $45$ Kilowatthours (kWh). We used this number to calculate that the same Raspberry Pi would have a power consumption of $1.4 mWs$. To set a high and unreachable upper boundary for $MT_3$, we set it to $10 mWs$ for the normalising all measurement results. The following example shows, how the previously calculated raw data measurement values of $X_{1MT_1}$, $X_{2MT_1}$, $X_{1MT_2c}$, $X_{1MT_3c}$ and $X_{2MT_3c}$  can be normalised using the \textit{MIN-MAX Normalisation} as described in (\ref{eq:7}):
\vspace{-5pt}
\begin{table}[H]
	\centering
	\begin{tabular}{lr}
		$\dot{X}_1 = 0 + \frac{(10ms - 0ms) * (1 - 0)}{1000 - 0}\>\>\>\>\>\>\>\quad\quad= 0.01$, & \\
		\rule{0pt}{15pt}$\dot{X}_2 = 0 + \frac{(20ms - 0ms) * (1 - 0)}{1000 - 0}\>\>\>\>\>\>\>\quad\quad= 0.02$, & \\
		\rule{0pt}{15pt}$\dot{X}_3 = 0 + \frac{(1.825\%- 0ms) * (1 - 0)}{1000 - 0}\>\>\>\>\>\quad\quad= 0.01825$, &$\qquad$as described in (\ref{eq:7})\\
		\rule{0pt}{15pt}$\dot{X}_4 = 0 + \frac{(0.005mWs - 0mWs) * (1 - 0)}{10 - 0}\quad= 0.0005$, & \\
		\rule{0pt}{15pt}$\dot{X}_5 = 0 + \frac{(0.010mWs - 0mWs) * (1 - 0)}{10 - 0}\quad= 0.0010$, & \\
	\end{tabular}
\end{table}

\vspace{-10pt}
After normalising each measurement, each normalised value has to be weighted using its calculated weight depending on the metric type. As described in (\ref{eq:6}) the weight is calculated by the number of metrics used divided by the total number of metrics. Since one representative metric is used for each metric type, the calculated weight can be described as $w_{MT_1} = w_{MT_2} = w_{MT_3} = \frac{1}{3}$. This means that each normalised measurement result has to be multiplied by $\frac{1}{3}$ before the aggregation. Finally, the following example shows how the \textit{Onion Layer Model} (\ref{eq:1}) can be used to calculate the \textit{Security Costs} of all \textit{security-related} task, performed by component $c$ during interaction $i$:
\vspace{-5pt}
\begin{table}[H]
	\centering
	\begin{tabular}{l}
		$SecurityCosts_{CPS_1} = (0.01 + 0.02 + 0.01825 + 0.0005 + 0.0010)*\frac{1}{3} = 0.01658\dot{3},$ as described in (\ref{eq:1})
	\end{tabular}
\end{table}

\vspace{-5pt}
The same approach as shown in Fig. \ref{fig:Fig3} and as explained in the examples above, was used to measure, normalise and aggregate the \textit{Functional Costs} and the \textit{Security Costs} of \textit{Use Case 1} and \textit{Use Case 2}. In this regard, each workload was executed separately and the SCMF was used to measure, normalise and aggregate the \textit{Total Costs} and \textit{Security Costs}. The \textit{Total Costs}, on the one hand, refer to the measurements of all tasks, no matter if they were \textit{functional} or \textit{security-related}. The \textit{Security Costs}, on the other hand, represent only the measurements of the \textit{security-related} tasks. The separation of the costs in those two groups enables to show the proportion of the \textit{Security Costs} in comparison to the \textit{Total Costs} for each workload. The following table summarises the measurement results of the experimental study for each workload regarding the \textit{Total Costs}:
\vspace{-5pt}
\begin{table}[H]
	\caption{Total Costs per Workload}
	\label{tab:res1}
	\begin{tabular}{ccccccccc}
		\toprule
		%\multicolumn{8}{c}{Total Costs}\\
		WL & Measurement & Min & Max & Median & Mean & Std.Dev. & Std.Err. & $\sum$\\
		\midrule
		\multirow{2}{*}{$WL_{1.1}$} & $\textless 25$\textdegree{}$C$ & $0.00083$ & $0.36000$ & $0.01800$ & $0.04377$ & $0.05060$ & $0.00221$ &  \multirow{2}{*}{$53.92297$}\\
		& $\textgreater 25$\textdegree{}$C$ & $0.00100$ & $0.28755$ & $0.02200$ & $0.05894$ & $0.07366$ & $0.00321$ &  \\\hline
		\multirow{2}{*}{$WL_{1.2}$} & $\textless 25$\textdegree{}$C$ & $0.00070$ & $0.34000$ & $0.02100$ & $0.04876$ & $0.05830$ & $0.00254$ & \multirow{2}{*}{$53.51435$}\\
		& $\textgreater 25$\textdegree{}$C$ & $0.00089$ & $0.39874$ & $0.02000$ & $0.05317$ & $0.06804$ & $0.00297$ & \\\hline
		\multirow{2}{*}{$WL_{2.1}$} & $\textless 25$\textdegree{}$C$ & $0$ & $0.08281$ & $0$ & $0.00472$ & $0.01431$ & $0.00062$ & \multirow{2}{*}{$36.15943$}\\
		& $\textgreater 25$\textdegree{}$C$ & $0.00100$ & $0.44075$ & $0.02642$ & $0.06415$ & $0.08218$ & $0.00359$ &  \\\hline
		\multirow{2}{*}{$WL_{2.2}$} & $\textless 25$\textdegree{}$C$ & $0$ & $0.10100$ & $0$ & $0.00480$ & $0.01471$ & $0.00064$ & \multirow{2}{*}{$31.57160$}\\
		& $\textgreater 25$\textdegree{}$C$ & $0.00100$ & $0.37200$ & $0.02066$ & $0.05534$ & $0.07307$ & $0.00319$ &  \\
		\bottomrule
	\end{tabular}
\end{table}

\vspace{-5pt}
The following table summarises the measurement results of the experimental study for each workload regarding the \textit{Security Costs}:
\vspace{-6pt}
\begin{table}[H]
	\caption{Security Costs per Workload}
	\label{tab:res2}
	\begin{tabular}{ccccccccc}
		\toprule
		WL & Measurement & Min & Max & Median & Mean & Std.Dev. & Std.Err. & $\sum$\\
		\midrule
		\multirow{2}{*}{$WL_{1.1}$} & $\textless 25$\textdegree{}$C$ & $0.00267$ & $0.13728$ & $0.01500$ & $0.04714$ & $0.05065$ & $0.00414$ & \multirow{2}{*}{$17.61541$}\\
		& $\textgreater 25$\textdegree{}$C$ & $0.00267$ & $0.27930$ & $0.02231$ & $0.07030$ & $0.08565$ & $0.00699$ & \\\hline
		\multirow{2}{*}{$WL_{1.2}$} & $\textless 25$\textdegree{}$C$ & $0.00244$ & $0.16359$ & $0.01400$ & $0.04861$ & $0.05337$ & $0.00436$ & \multirow{2}{*}{$16.08108$}\\
		& $\textgreater 25$\textdegree{}$C$ & $0.00178$ & $0.23166$ & $0.01500$ & $0.05859$ & $0.07342$ & $0.00599$ & \\\hline
		\multirow{2}{*}{$WL_{2.1}$} & $\textless 25$\textdegree{}$C$ & $0$ & $0$ & $0$ & $0$ & $0$ & $0$ & \multirow{2}{*}{$10.97640$}\\
		& $\textgreater 25$\textdegree{}$C$ & $0.00267$ & $0.29200$ & $0.02486$ & $0.07318$ & $0.08504$ & $0.00694$ & \\\hline
		\multirow{2}{*}{$WL_{2.2}$} & $\textless 25$\textdegree{}$C$ & $0$ & $0$ & $0$ & $0$ & $0$ & $0$ & \multirow{2}{*}{$10.02343$}\\
		& $\textgreater 25$\textdegree{}$C$ & $0.00200$ & $0.31474$ & $0.02239$ & $0.06682$ & $0.08074$ & $0.00659$ & \\
		\bottomrule
	\end{tabular}
\end{table}

\section{\textbf{Discussion and Future Work}}\label{section:discussion}

The results of the experimental study show that changing the component that initiates the \textit{Closed-Loop Temperature Control} interaction can have significant impact on the overall performance. To obtain these measurement results, we followed the steps of the SCMF to evaluate the performance of each workload from Table \ref{tab:workloads} by using the metrics from Table \ref{tab:metrics}. As described in Section \ref{subsection:SCMF} the steps of the SCMF can be summarised as follows:
\begin{itemize}
	\item Step 1: perform \textit{Algorithmic Complexity Analysis ($MT_0$)}, as described in (\ref{eq:2})
	\item Step 2: perform all \textit{Difference Measurements ($MT_1$)}, \textit{Sample Measurements ($MT_2$)} and \textit{Overall Result Measurements ($MT_3$)} for each workload, as described in (\ref{eq:3}), (\ref{eq:4}) and (\ref{eq:5})
	\item Step 3: normalise each raw data measurement result from Step 2, as described in (\ref{eq:6})
	\item Step 4: calculate the weights for each metric type, as described in (\ref{eq:7})
	\item Step 5: aggregate each normalised result from Step 3 multiplied by the corresponding calculated weight from Step 4, as described in (\ref{eq:1})
	\item Step 6: add the result from Step 1 as a constant to the overall result from Step 5
\end{itemize}

The results in Table \ref{tab:res1} and Table \ref{tab:res2} show that \textit{Use Case 2} performs around 33\% better compared to \textit{Use Case 1}. The main reason for this is that in \textit{Use Case 1} all components participate during the interaction, no matter if the measured room temperature is above the predefined limit or not. This means that each component executes its tasks, even if the air-conditioning system does not have to be activated in the end. In \textit{Use Case 2}, however, the entire \textit{Closed-Loop Temperature Control} interaction including all its components is only executed when the measured temperature limit is above the limit. The other times, when the temperature is below the limit, only the component with the temperature sensor ($C_2$) is involved while the others are idle. This idle state of components is mainly shown for $WL_{2.1}$ and $WL_{2.2}$, since these workloads are executed mainly on the \textit{Use Case 2}.

The results also show that both use cases spend around 30\% of their time computing resources and power for performing security-related tasks. Having 30\% of tasks \textit{security-related} is the same in both use cases and in each workload. The reason for this is that the \textit{security-related} tasks are always executed in the same way, no matter if the temperature limit has been reached or not. So once an entire interaction cycle where all components are involved is executed the number of \textit{security-related} tasks stays the same. This means that the performance improvement of 33\% is mainly influenced by the temperature measurement and whether the predefined limit has been reached or not. Nevertheless, this simple experiment shows that the \textit{Total Costs} of an interaction can be improved significantly by changing a small parameter (e.g. the initiating component from $C_1$ to $C_2$). As a result, the \textit{Security Costs} also improved as a consequence of the reduction in \textit{Total Costs}.

The results also confirm that the HTTPS protocol produces additional costs in comparison to its insecure counterpart (HTTP). Even though this additional overhead is small, it has to be considered when aggregating the overall \textit{Security Costs} of a CPS. In this case the difference between $WL_{1.1}$ and $WL_{1.2}$ has to be added to the overall \textit{Security Costs} of \textit{Use Case 1}. The same has to be done for $WL_{2.1}$ and $WL_{2.2}$ in \textit{Use Case 2}. Additionally, the algorithmic time complexity estimation constant has to be added to the total sum to finally calculate the overall \textit{Security Costs} for both use cases:
\begin{itemize}
	\item[]\textit{Use Case 1:}\space\space\space$Security Costs_{CPS_1} = (53.92297 - 53.51435) + 17.61541 = 18.02403 + f_{MT_0}(n)$
	\item[]\textit{Use Case 2:}\space\space\space$Security Costs_{CPS_1} = (36.15943 - 31.57160) + 10.97640 = 15.56423 + f_{MT_0}(n)$
\end{itemize}

Since CPS are large-scaled, distributed systems that can  become very complex it is getting increasingly difficult to keep track, monitor and understand the \textit{behaviour} of its components. Developing these systems poses a huge challenge since they have to be designed in a way that ensures security throughout all distributed computing nodes without impacting the overall performance. The results of the \hyperref[section:experiment]{\textit{Experimental Study}} showed that the overall performance of a CPS could be significantly improved by designing the interaction to execute all \textit{security-related} tasks only when necessary. Or in other words, by deciding to implement an interaction in a certain way can have a positive/negative impact on the performance of a CPS. The presented SCMF can be used to measure the performance of a CPS by using different metrics, normalise these results to create a generic \textit{Cost Unit} and aggregate them using the \textit{Onion Layer Model}. In further work we would like to extend the SCMF as shown in \hyperref[fig:Fig3]{Fig. \ref{fig:Fig3}} by researching performance measurement metrics and their efficacy in more detail. In this regard, our plan is to identify all relevant metrics for measuring the performance of CPS, catagorise them by metric types and evaluate their efficacy for measuring \textit{Security Costs}. Additionally, we are planning to also search the literature for all techniques for scaling measurement results with different units to a generic \textit{Cost Unit} and discuss their trade-offs. These two additions would add more depth to the SCMF by providing a \textit{Metric Catalogue} for measuring \textit{Security Costs} and relevant scaling techniques including their trade-offs. 

Another interesting application field for the SCMF is to use it for optimising the performance of interactions at \textit{Design Time}. In this regard all measured and scaled results could be used to find an optimal solution for an interaction which achieves the same goal while producing the least \textit{Security Costs}. More precisely, as shown in the \hyperref[section:experiment]{\textit{Experimental Study}} we defined different workloads for the same interaction that were measured, analysed and compared. For this we used different types (secure and insecure) of the same communication protocol to identify how much the "S" costs in HTTPS. The same approach could be used to define workloads where one interactions uses many different secure protocols to identify which of them produces the least \textit{Security Costs}. We also plan to investigate methodologies for trying different configurations of an interaction (e.g. different secure protocols) and measure their resulting \textit{Security Costs}. 

\section{\textbf{Conclusion}}\label{section:conclusion}

In this article we propose a framework for measuring, normalising and aggregating \textit{Functional Costs} and \textit{Security Costs} of CPS. The SCMF provides an \textit{Onion Layer Model} for formally describing a CPS based on its interactions, their participating components and performed \textit{functional} and \textit{security-related} tasks. In addition to that we present four \textit{Metric Types} ($MT_0$, $MT_1$, $MT_2$, $MT_3$) and show how they can be used to measure these tasks and calculate the resulting costs. Furthermore, we show how the measured and calculated raw data values can be normalised and weighted. This enables the transformation of  measurement values with different units to a common \textit{Cost Unit} and to aggregate the \textit{Functional Costs} and \textit{Security Costs} using the presented \textit{Cost Aggregation} algorithm.

In addition to the model, we also describe the minimum requirements to set up and measure the performance of a CPS. The presented architectural design describes the need for an IoT-Framework (such as Arrowhead Framework) to draw the system boundaries of a CPS and control the behaviour within it. Furthermore, we demonstrate that the \textit{Functional Costs} and the \textit{Security Costs} of a CPS can be measured by using an agent-based bytecode instrumentation tool (e.g. Pinpoint APM) in combination with the PiLogger One board. Based on the model and architectural design we present our experimental testbed setup of a CPS with a \textit{Closed-Loop Temperature Control} interaction. Finally, we describe two ways of implementing this interaction (\textit{Use Case 1}, \textit{Use Case 2}), measure their overall performance and show which implementation is more efficient. Furthermore, by extracting the \textit{Security Costs} from the \textit{Total Costs} we show the portion of the overhead imposed by executing \textit{security-related} tasks. This can also be helpful at \textit{Design Time} to evaluate which implementation of an interaction preserves its cyber-security countermeasures while optimally using its resources.

Our experimental study reveals that the \textit{Use Case 2} produced around $33\%$ less \textit{Security Costs} compared to \textit{Use Case 1}, while executing the \textit{Closed-Loop Temperature Control} interaction. Moreover, the results confirmed that the "S" in HTTPS produces additional costs which have to be included in the overall result of the aggregated \textit{Security Costs} for both use cases. Summarising, in this article we presented the SCMF and described each step of measuring, normalising and aggregating the \textit{Functional Costs} and the \textit{Security Costs} of CPS. In addition to that we also described the architectural elements needed to be able to apply the SCMF in a real use case. Finally, we showed in an experimental study that the SCMF can be used to model the overall performance (\textit{Total Costs}) of two interactions to identify which of them is more efficient.

\begin{acks}
	We would like to thank the anonymous reviewers for their constructive and helpful comments to improve this article. 
	Special thanks to Helmut Peter for his electrotechnical expertise during the experimental setup phase.
	The research has been carried out in the context of the project MIT 4.0 (FE02), funded by IWB-EFRE 2014 – 2020 and coordinated by Forschung Burgenland GmbH.
\end{acks}

\bibliographystyle{ACM-Reference-Format}
\bibliography{15_references}

\end{document}